\documentclass[sort&compress,square, numbers,sort,nofootinbib]{revtex4-2}

\usepackage[utf8]{inputenc}
\usepackage{url} 
\usepackage{float}
\usepackage{amsmath, amssymb}
\usepackage{graphicx}
\usepackage{caption}
\usepackage{subcaption}

\usepackage{array}
\usepackage[table]{xcolor}
\definecolor{LightCyan}{rgb}{0.88,1,1}

\usepackage{tikz}

\usetikzlibrary{automata,positioning, arrows.meta, quotes, backgrounds,decorations.markings,decorations.pathreplacing}

\begin{document}

\title{Proca in an Expanding Universe}
	
\author{Shaun David Brocus Fell}
\email{fell@thphys.uni-heidelberg.de}
\affiliation{Institute for Theoretical Physics, Universit{\"a}t Heidelberg , Philosophenweg 16, 69120 Heidelberg, Germany}

\author{Lavinia Heisenberg}
\email{heisenberg@thphys.uni-heidelberg.de}
\affiliation{Institute for Theoretical Physics, Universit{\"a}t Heidelberg , Philosophenweg 16, 69120 Heidelberg, Germany}

\begin{abstract}
The superradiant growth of massive vector fields in rotating black hole spacetimes has garnered significant attention in recent literature. However, the majority of these studies overlook the influence of a cosmological constant, which likely constitutes the primary energy content of our universe. In this paper, we extend recent research by incorporating a cosmological constant into the Einstein+Proca system and numerically evolving the resulting equations of motion. Utilizing the newly released GRBoondi numerical relativity code, designed specifically for the numerical evolution of (generalized) Proca fields, we discover that parameters causing a growing instability in the $\Lambda=0$ scenario transition to a decaying state when $\Lambda>0$. This results in a more intriguing phenomenology. These simulations pave the way for future full Einstein+Proca simulations to explore the secular decay of the resultant cloud from gravitational emission.
\end{abstract}

\maketitle

\section{Introduction}
As an effective field theory, General Relativity performs exceptionally well at intermediate scales. However, it faces significant challenges at both extremely small and vast cosmological scales, necessitating extensions to the theory \cite{Heisenberg:2018vsk}. Its problematic reconciliation with quantum mechanics and the inconsistencies observed in cosmology indicate the need for new physics. Two of the most pressing questions in contemporary cosmological research pertain to the seemingly invisible components of the universe, whose existence is inferred from compelling observational data \cite{planck2020, riess2023crowded, ghirardini2024srgerosita}. These unknown components are colloqiually dubbed 'dark matter' and 'dark energy', the latter being responsible for the accelerated expansion of the universe. The former is predicted to exist from observations of galactic rotation curves and gravitational lensing measurements \cite{2009GReGr..41..207Z,1937ApJ....86..217Z, 1936ApJ....83...23S, 1939LicOB..19...41B, 1970ApJ...159..379R, 1975ApJ...201..327R, 1940ApJ....91..273O, 1985ApJ...289...81R, 1988MNRAS.234..131P, einsteinrelativity}. The nature of these components remains elusive, however there are many experiments and theoretical analyses today that are looking to uncover these invisible sectors of our model of the universe (see for example \cite{Aprile_2023, Bartram_2021, Jungman1996, Preskill1983, essig2013dark, Holdom1986, Galison1984} and references therein).

The dark photon in particular is a well-motivated candidate for dark matter and it has several possible production mechanisms, including a misalignment mechanism \cite{Nelson_Scholtz_2011}, arising naturally in certain string theories \cite{Goodsell_Jaeckel_Redondo_Ringwald_2009}, tachyonic couplings to misaligned axions \cite{Co_2019}, and quantum flucuations during the inflationary epoch \cite{Graham_2016}.  Constraints on the couplings of such particles to the Standard Model come from equivalence principle tests, such as of the Eöt-Wash group~\cite{PhysRevD.50.3614, Schlamminger_2008} and Lunar Laser Ranging groups~\cite{PhysRevLett.93.261101, TURYSHEV_2007}. Near future constraints will come from GW measurements, for example from black hole superradiance~\cite{Baryakhtar_Lasenby_Teo_2017, East_Pretorius_2017, East_2017, Pierce_Riles_Zhao_2018}. The dark photon does not need to coincide with the photon of the Standard Model. Observational data strongly suggest that it must be a completely new particle beyond the Standard Model. This dark photon could exhibit intriguing interactions, including derivative self-couplings \cite{Heisenberg:2014rta}. This could lead to fascinating cosmological applications beyond just explaining dark matter \cite{DeFelice:2016yws,DeFelice:2016uil}.

Regarding dark photon applications as dark matter, superradiance will soon offer stringent tests on the mass of massive vector fields, thereby constraining the allowable parameter space for dark matter candidates. Due to this, several theoretical analyses have been carried out to understand the evolutionary behavior of a massive vector field surrounding black holes \cite{Baryakhtar_Lasenby_Teo_2017, Brito_2015, Cardoso_2018, Siemonsen_East_2020, East_2018, East_2017, PhysRevD.108.083010}. It has been shown that the Proca field exhibits a very rapid instability phase before saturation, afterwhich it experiences a secular decay due to gravitational radiation emission. The instability phase can be very rapid, typically on the order of seconds. On the other hand, the secular decay due to gravitational radiation emission can last from months to several billion years depending on the mass of the vector field. Both the fast and slow gravitational radiation decay can be used to constrain the vector field mass using upcoming gravitational wave (GW) antennas, such as the LISA mission \cite{lisa1, lisa2}.

It is thus important to develop accurate models of the evolution of the Proca cloud surrounding black holes. Previous studies have focused on the evolution of the Proca cloud surrounding a Schwarzschild or Kerr black hole, with no influence from the cosmic expansion \cite{Baumann_Chia_Stout_ter, East_2017, East_2018, Rosa_Dolan_2012, FKKS_2018, Herdeiro_Radu_Runarsson_2016, Dolan_2018, Siemonsen_East_2020, Cutler_Kennefick_Poisson_1994, Yoshino_Kodama_2014, Santos_Benone_Crispino_Herdeiro_Radu_2020, Baryakhtar_Lasenby_Teo_2017,PhysRevD.108.083010}. This is normally a justifiable approximation since recent cosmological observations have shown that the cosmological constant is quite small \cite{planck2020, riess2023crowded, ghirardini2024srgerosita}. Nonetheless, the effects of a cosmological constant on the growth rate of the Proca cloud is still required to complete the understanding of Proca superradiance in our universe. It is thus the goal of this study to further the understanding of Proca in a Kerr-de Sitter background. We opt to neglect the backreaction of the Proca cloud on the background since the magnitude of the stress-energy tensor of the Proca field is negligible. This is a common assumption for the initial phase of superradiant instability studies since the cloud begins as a perturbation on the spacetime. As long as the Proca field does not constitute a dominant contribution and its interactions remain weak, the backreaction effect is expected to be minimal. Going beyond the initial phase to studies of the resulting gravitational radiation would require full mutual evolution of the matter and metric variables, hence this is beyond the scope of this study and is relegated to future work.

The paper is divided into 4 sections. Section~\ref{sec:theory} takes care of outlining the basic theoretical background, including the equations of motion and the particular coordinate system. Section~\ref{sec:results} then discusses the results of this study. Finally, Section~\ref{sec:conclusion} concludes and discusses future work. An appendix expands on the various formulas utilized in the main text.

\section{Theory} \label{sec:theory}
The starting point is the Einstein-Hilbert-Proca theory, where the Einstein-Hilbert action is augmented by the Proca action
\begin{equation}
S[g_{\mu \nu},A^{\sigma}] = \int d^4x \sqrt{-g} \left(\frac{1}{16 \pi G} (\mathfrak{R} - 2 \Lambda) - \frac{1}{4} F^{\mu \nu} F_{\mu \nu} - \frac{1}{2} \mu^2 A^{\mu}A_{\mu} \right),
\end{equation}
where $g_{\mu \nu}$ is the metric, $g$ its determinant, $\mathfrak{R}$ is the Ricci scalar, $\Lambda$ is the cosmological constant, and $A^{\mu}$ is the Proca 4-vector. Variation of $S[g,A]$ with respect to these two fields yields the equations of motion
\begin{align} \label{eq:einsteinproca}
		G^{\rho \sigma} + \Lambda g^{\rho \sigma} &= 8 \pi \left( \frac{-1}{4} F_{\mu \nu}F^{\mu \nu} g^{\rho \sigma} + F^{\rho \nu} F^{\sigma}_{\nu} - \frac{1}{2} \mu^2 g^{\rho \sigma} A_{\mu} A^{\mu} +  \mu^2 A^{\rho} A^{\sigma} \right) = 8 \pi \mathfrak{T}_{\mu \nu}\\
		0&= \nabla_{\rho} F^{\rho \sigma} - \mu^2 A^{\sigma}\;,
\end{align}
where $G^{\rho \sigma}$ is the Einstein tensor and $\mathfrak{T}_{\mu \nu}$ is the stress energy tensor of the Proca field. We assume the backreaction of the Proca field on the spacetime is negligible, as is usually the case, meaning we can set $\mathfrak{T}_{\mu \nu}=0$ without significant loss of accuracy. This implies that the field equations reduce to 
\begin{align} \label{eq:einsteinprocared}
		G^{\rho \sigma} + \Lambda g^{\rho \sigma} &= 0\\
		  \nabla_{\rho} F^{\rho \sigma} - \mu^2 A^{\sigma} &= 0\;. \label{eq:procaeq}
\end{align}
The solution of the gravity sector describing a spinning black hole is called the Kerr-de Sitter (KdS) solution. This solution is in fact a special case of the more general Plebanski-Demianski family of metrics \cite{Griffiths_Podolsky_2009, PLEBANSKI197698}. We choose to work in the Kerr-Schild form of the solution, which takes the form
\begin{equation} \label{eq:KdS}
g_{\mu \nu} = g_{0, \mu \nu} + 2 H K_{\mu} K_{\nu}\;,
\end{equation}
where $g_{0, \mu \nu}$ is the background de Sitter metric and $K_{\mu}$ is a null vector (with respect to both $g$ and $g_{0}$). In Kerr-Schild coordinates $(t, r, \theta, \phi)$, the de Sitter background metric takes the form
\begin{equation} \label{eq:fullmetric}
g_{0,\mu \nu} = 
\begin{bmatrix}
\frac{-\Delta_{\theta}}{\Theta} \Lambda_r & 0 & 0 & 0 \\
0 & \frac{\rho^2}{(r^2 + a^2)} \Lambda_r & 0 & 0 \\
0 & 0 & \frac{\rho^2}{\Delta_{\theta}} & 0 \\
0 & 0 & 0 & \frac{(r^2 + a^2)}{\Theta} \sin{\theta}
\end{bmatrix}\;,
\end{equation}
where we've defined
\begin{equation}
\begin{split}
\Delta_{\theta} &= 1 + \frac{\Lambda}{3} a^2 \cos{\theta}^2 \\
\Theta &= 1 + \frac{\Lambda}{3} a^2
\end{split}
\quad \quad
\begin{split}
\Delta_r &= r^2 - 2Mr + a^2 - \frac{\Lambda}{3} r^2 (r^2 + a^2) \\
\rho^2 &= r^2 + a^2 \cos{\theta}^2 
\end{split}
\end{equation}
and $H = \frac{2Mr}{\rho^2}$. The $\Delta_r$ definition becomes important for the analysis of the black hole horizons. The poles of $\Delta_r$ correspond to the poles of the metric in Boyer-Lindquist coordinates \cite{Akcay_Matzner_2011}.
In a similar fashion, the null vector is defined via 
\begin{equation}
K_{\mu} = \left( \frac{\Delta_{\theta}}{\Theta}, \frac{\rho^2}{(r^2 + a^2)\Lambda_r}, 0, \frac{-a}{\Theta} \sin^2(\theta) 
\right)
\end{equation}

Prior to plugging this metric into the numerical solver, we need to understand the basic causal structure of the background spacetime on which the Proca cloud will evolve. This amounts to determining the location of the horizons, which follows by solving the quartic polynomial $\Delta_r = 0$. The \emph{existence} of the horizons can be determined by analyzing the discriminant of the quartic polynomial, which will be denoted by $Q$. The positivity of $Q$ guarantees that either all four roots are simultaneously either real or complex. This means we only have to verify one of the roots is real to ensure the other 3 are as well. The discriminant is easily solved for and takes the form
\begin{align}
Q =& -\frac{16 \Lambda}{243} \left[ 12 a^8 \Lambda^8 + a^{10} \Lambda^{4} + a^2 \left( 81 - 891 M^2 \Lambda \right) + 3 a^6 \Lambda^2 \left(18 + M^2 \Lambda \right) \right. \\
& \left.  + 81 M^2 \left(-1 + 9 M^2 \Lambda \right) + 27 a^4 \Lambda \left( 4 + 11 M^2 \Lambda \right) \right]\;.
\end{align}

\begin{figure*}
    \centering
    \includegraphics[width=0.5\linewidth]{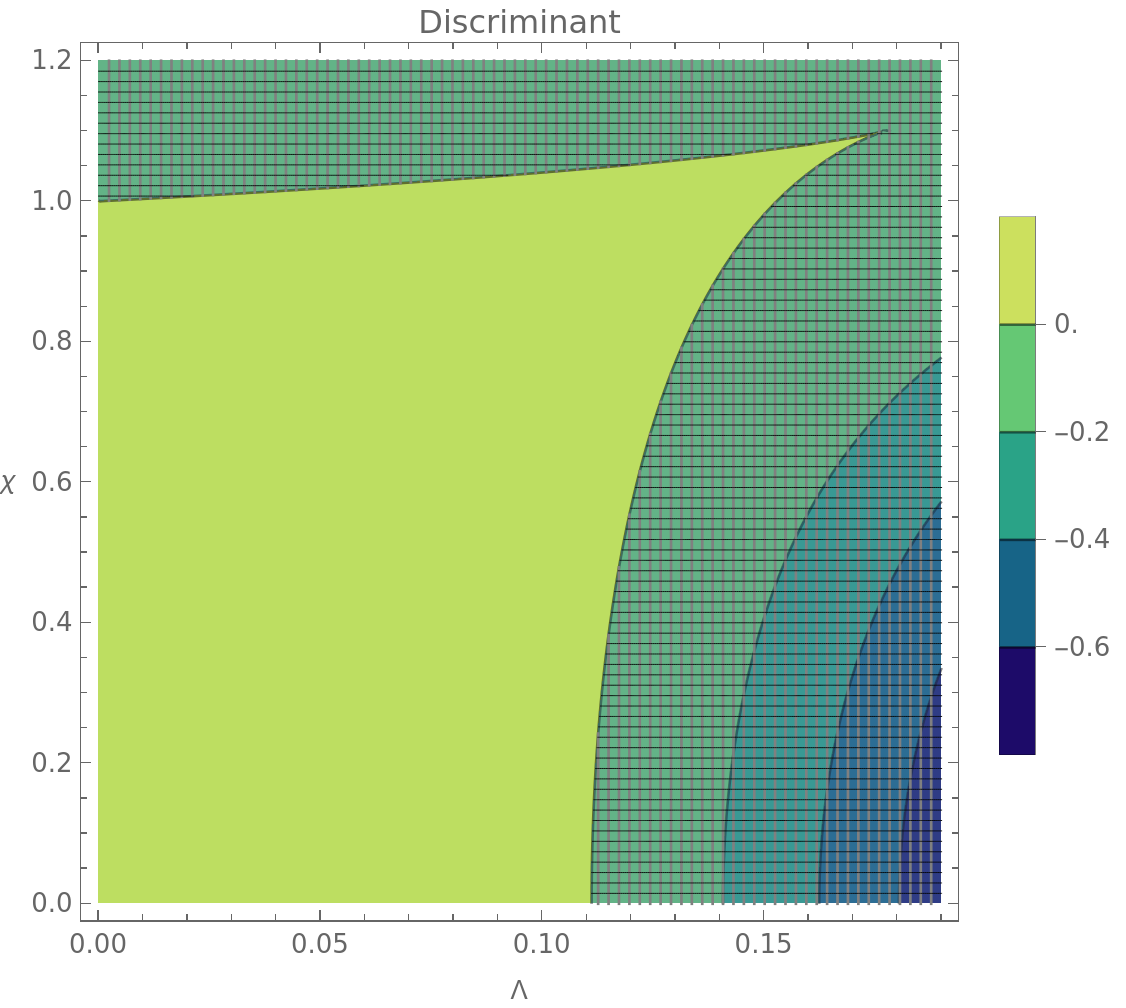}
    \caption{Plot of the values of the discriminant $Q$. The shaded region shows the unallowed parameters for the existence of a Kerr-de Sitter black hole. The unshaded region are the allowable parameters.}
    \label{fig:discriminant}
\end{figure*}

Fig.~\ref{fig:discriminant} shows a contour plot of $Q$, with the spin and cosmological constant rescaled by the black hole mass. The region $Q>0$ shows the allowed parameters for the background. The shaded region denotes the disallowed region. Beyond the allowed region, one or more horizons will disappear, yielding a naked singularity. Fig.~\ref{fig:discriminant} hence tells us the allowed values of the black hole parameters that can be used in the simulations. An absolute maximum value of the cosmological constant is $\Lambda_{max} = \frac{16}{45 + 26 \sqrt{3}} \frac{1}{M^2}$ and corresponding absolute maximum value of the black hole spin of $a_{max}=\sqrt{\frac{9}{16} + \frac{3\sqrt{3}}{8}}M$. This point corresponds to the cusp of the non-shaded region on the upper-right quadrant of Fig.~\ref{fig:discriminant}.

Another interesting feature is the existence of a minimum spin for a certain range of values for the cosmological constant. For spacetimes satisfying $\Lambda \geq 1/9$, the black hole is required to possess spin in order for the Kerr-de Sitter black hole to exist. That is to say, for cosmological constant values greater than $1/9$, static black holes in the form of Eq.~\ref{eq:KdS} do not exist.

We label the three positive roots as $r_{-}$, $r_{+}$, and $r_{\Lambda}$, which denote the inner, outer, and cosmological horizons, respectively. The fourth root is negative and corresponds to a 'horizon' inside the singularity at $r=0$. In Kerr-Schild coordinates, besides the singularity at $r=0$, the poles in the metric disappear, including $r_{\Lambda}$. However, a new pole appears at $\tilde{r}_{\Lambda}=\sqrt{\frac{3}{\Lambda}}$. We find that $r_{\Lambda} \leq \tilde{r}_{\Lambda}$ throughout the allowed parameter space. In the limit of small cosmological constant, 
\begin{equation}
r_{\Lambda} = \tilde{r}_{\Lambda} - 1 + \mathcal{O}\left( \sqrt{\Lambda} \right)\;.
\end{equation}

\begin{figure*}
    \centering
    \includegraphics[width=0.5\linewidth]{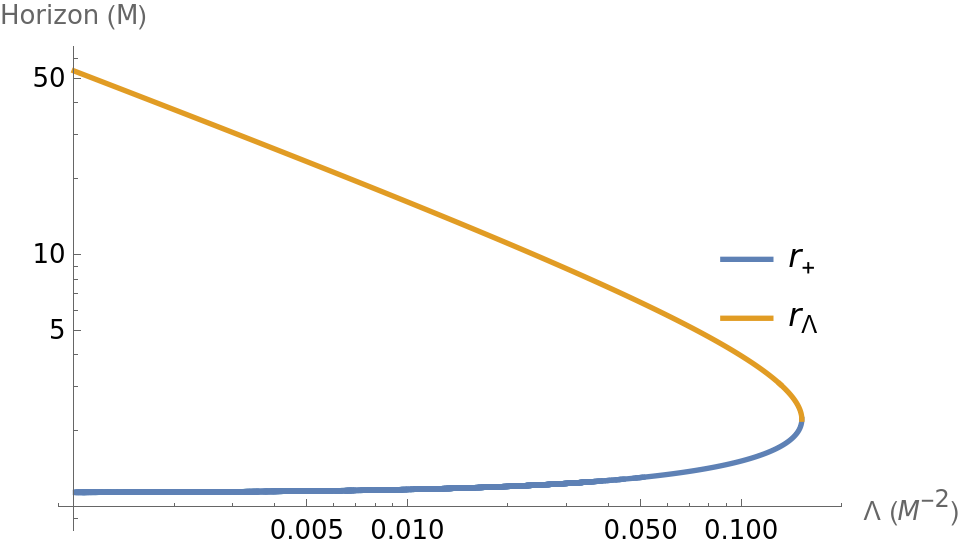}
    \caption{Evolution of the cosmological and outer horizons, eventually merging. A value of $a=0.5$ is chosen as a representative value. Past the value of $\Lambda$ where the horizons merge, no Kerr-de Sitter black hole is possible.}
    \label{fig:rprc}
\end{figure*}

A peculiar property of the causal structure is the existence of a maximum cosmological constant which permits the existence of a black hole. At the maximum cosmological constant value $\Lambda_{max}$, the outer and cosmological horizons merge, see for example Fig.~\ref{fig:rprc}. Additionally, dimensionless spin values greater than unity are allowed without producing a naked singularity, as long as the cosmological constant is greater than zero.

On the other hand, the solution in the Proca sector of Eqs.~\ref{eq:einsteinprocared} is unclear, since analytic studies have yet to be performed. However, based on knowledge from the Proca on Kerr solutions, we can expect the cloud to undergo a superradiant instability due to the existence of the black hole horizon.  This implies that the energy of the cloud will be of the form
\begin{equation}
E(t) = E_0 e^{2 i \omega t}\;,
\end{equation}
where $\omega$ is the complex frequency of the Proca field. The factor of 2 comes from the fact that the stress-energy tensor of the Proca field is quadratic in the Proca field itself. Since the frequency is of the form $\omega = \omega_r + i \omega_i$, the imaginary part implies an exponential evolution superimposed over an oscillatory one. The imaginary frequency is what is solved for in this paper.

\subsection{3+1 decomposition}
Towards a numerical solution of the Proca system, the next step is to decompose the field equations into a form pertinent for numerical computations. We follow the standard procedure, which is to decompose the spacetime via a foliation into a series of three dimensional time-like hypersurfaces. Since the spacetime admits a time-like killing vector, we can choose the Kerr-Schild time coordinate as the function that defines the foliation leaves. Hence, the $(r,\theta,\phi)$ coordinates become coordinates on the hypersurfaces. We thus define our $3+1$ decomposition as \cite{Gourgoulhon2012}
\begin{align}
\alpha &= - \frac{g}{\gamma} \\
\beta^i &= - \frac{g^{0i}}{g^{00}}\\
\gamma_{ij}&= g_{ij}
\end{align}
where $g$ and $\gamma$ are the determinants of the full and spatial metrics, respectively, $\alpha$ is the lapse function, $\beta^i$ is the shift vector, $\gamma_{ij}$ is the spatial metric, and Latin indices range from one to three. In the coordinate system of Eq.~\ref{eq:fullmetric}, these take the form
\begin{align}
\alpha &= \sqrt{ \frac{ \Delta_{\theta} \Lambda_r^2 \rho^2}{\Gamma}} \\
\beta^i &= 
\begin{bmatrix}
\frac{ 2Mr(a^2 + r^2) \Delta_{\theta} \Lambda_r}{2a^2Mr\sin^2(\theta) \Lambda_r + \Theta (2Mr + (a^2 + r^2)\Lambda_r)\rho^2}\;\;\;\; & 0\;\;\;\; & -\frac{ 2Mra \Delta_{\theta} \Lambda_r}{2a^2Mr\sin^2(\theta) \Lambda_r + \Theta (2Mr + (a^2 + r^2)\Lambda_r)\rho^2} 
\end{bmatrix} \\
\gamma_{ij} &= 
\begin{bmatrix}
\frac{\rho^2 (2Mr + (a^2 + r^2) \Lambda_r)}{\Lambda_r^2 (a^2 + r^2)^2}\;\;\;\; & 0\;\;\;\; & -\frac{2aMr\sin^2(\theta)}{a^2 \Theta \Lambda_r + r^2 \Theta \Lambda_r} \\
0\;\;\;\; & \frac{\rho^2}{\Delta_{\theta}} \;\;\;\;& 0\\
-\frac{2aMr\sin^2(\theta)}{a^2 \Theta \Lambda_r + r^2 \Theta \Lambda_r}\;\;\;\; & 0\;\;\;\; & \frac{\sin^2(\theta)}{\Theta^2} \left( \Theta (r^2 + a^2) + \frac{2a^2Mr\sin^2(\theta)}{\rho^2} \right)
\end{bmatrix}\;,
\end{align}
where we've defined $\Gamma = 2Mr \Delta_{\theta} + \rho^2 \Theta \Lambda_r$.
As our numerical solver computes the time evolution using cubic cells, we transform the $(r,\theta,\phi)$ coordinates to a Cartesian-like coordinate system defined by 
\begin{align}
x &= r \sin (\theta) \cos(\phi) \label{eq:cartx} \\
y &= r \sin (\theta) \sin(\phi) \label{eq:carty} \\
z &= r \cos (\theta)\;. \label{eq:cartz}
\end{align}
The shift, spatial metric, and all derivatives are then transformed using the resulting Jacobian matrix. See App.~\ref{app:coordtransform} for more details.
The last step is to decompose the Proca equations Eq.~\ref{eq:procaeq} under the foliation. A standard calculation yields
\begin{align} \label{eq:procadecomp1}
\frac{1}{\alpha} \mathcal{L}_m E^i &= E^i K - D^i Z + \mu^2 X^i - \frac{2}{\alpha}D_j \left(\alpha D^{[j}X^{i]}\right) \\
\frac{1}{\alpha} \mathcal{L}_m Z &= -\mu^2  \phi - D_i E^i - \kappa Z \\
\frac{1}{\alpha} \mathcal{L}_m X_{i} &= -E_i - D_i \phi - \phi D_i ln(\alpha) \\
\frac{1}{\alpha} \mathcal{L}_m \phi &=  \frac{Z}{\mu^2} +  \phi K - D_i X^i - x^{i}D_i ln(\alpha)\;,  \label{eq:procadecomp2} 
\end{align}
where $E^i = \gamma^i_{\mu} F^{\mu \nu}$, $X_i = \gamma_{i \mu} A^{\mu}$, $\phi = -n_{\mu}A^{\mu}$, Z is an auxiliary field introduced to damp violations of the Proca constraint with a tuning parameter $\kappa$ \cite{East_2017, Clough_2022, Zilh_o_2015}, and $n^{\mu}$ is the time-like normal to the spatial hypersurfaces.

\section{Numerical results} \label{sec:results}
For the numerical computation of the Proca field, we use the recently released GRBoondi software \cite{fell2024grboondi}. GRBoondi is an offshoot of the GRChombo \cite{Andrade2021} framework that is catered towards numerical solutions of Generalized Proca theories \cite{Heisenberg:2014rta}. GRBoondi allows for the numerical evolution of Generalized Proca theories on arbitrary fixed backgrounds and so is an excellent choice for the system under consideration here, even if we concentrate here solely on the mass term. Generalizing our study to the more intriguing case of self-interacting Generalized Proca fields is straightforward with GRBoondi. 

GRBoondi solves the Proca equations Eq.~\ref{eq:procadecomp1}-\ref{eq:procadecomp2} using an adaptive mesh refinement (AMR) grid with time evolution determined using a Runge-Kutta fourth-order time integrator and fourth-order accurate finite difference stencils for the spatial derivatives.  We use a box width of $L=60M$ with $N=192$ grid points across each edge of the computational box. We use 4 refinements levels at a $2:1$ refinement ratio, resulting in a resolution of the finest level of $dx_{fine} = 0.01953M$. To prevent boundary effects from contaminating the simulation, we use Sommerfeld-outgoing radiation boundary conditions, which allows oscillations to exit the simulation region with minimal reflections due to finite-size effects. This is especially important since we introduced an auxiliary field which dampens violations of the constraint equation and the evolution equation for $Z$ is a generalized telegraph equation. This implies that not only are the values of $Z$ damped, but also propagate at the speed of light. Hence, the outgoing radiation boundary conditions are vital for ensuring violations of the constraint equation propagate outside the computational domain.

To understand the effect of a cosmological constant on the dynamical evolution of the superradiant Proca cloud, we perform a various number of simulations with parameters that yield the highest growth rates. We choose three different values of the cosmological constant, $\Lambda = \left(5 \cdot 10^{-6}, 10^{-4}, 10^{-3} \right)$. Higher values of the cosmological constant are more difficult to simulate numerically as the cosmological horizon quickly becomes small. We reserve probing this region of the parameter space to future studies, which will likely entail a new coordinate system. Additionally, we fix the black hole spin to $\chi = 0.99$. We sample the Proca mass at 6 different values, $\mu = \left( 0.35, 0.4, 0.45, 0.5, 0.6, 0.7 \right)$. In addition to the main simulations, we also perform a convergence study to ensure our choice of resolution produces accurate data, which we discuss in sec.~\ref{app:convertest}.

For initial data, we take a Gaussian profile with width determined by analytic approximation studies \cite{PhysRevD.108.083010}, $r_0 = \frac{1}{M^2 \mu}$. The initial data is then $A_x = \frac{A}{\gamma} e^{-\frac{r}{r_0}} $, where $A$ is some pre-determined amplitude which we take to be $A=0.1$, and all other variables are chosen to be zero. 

\begin{figure}[hbtp]
    \centering
    \includegraphics[width=1\linewidth]{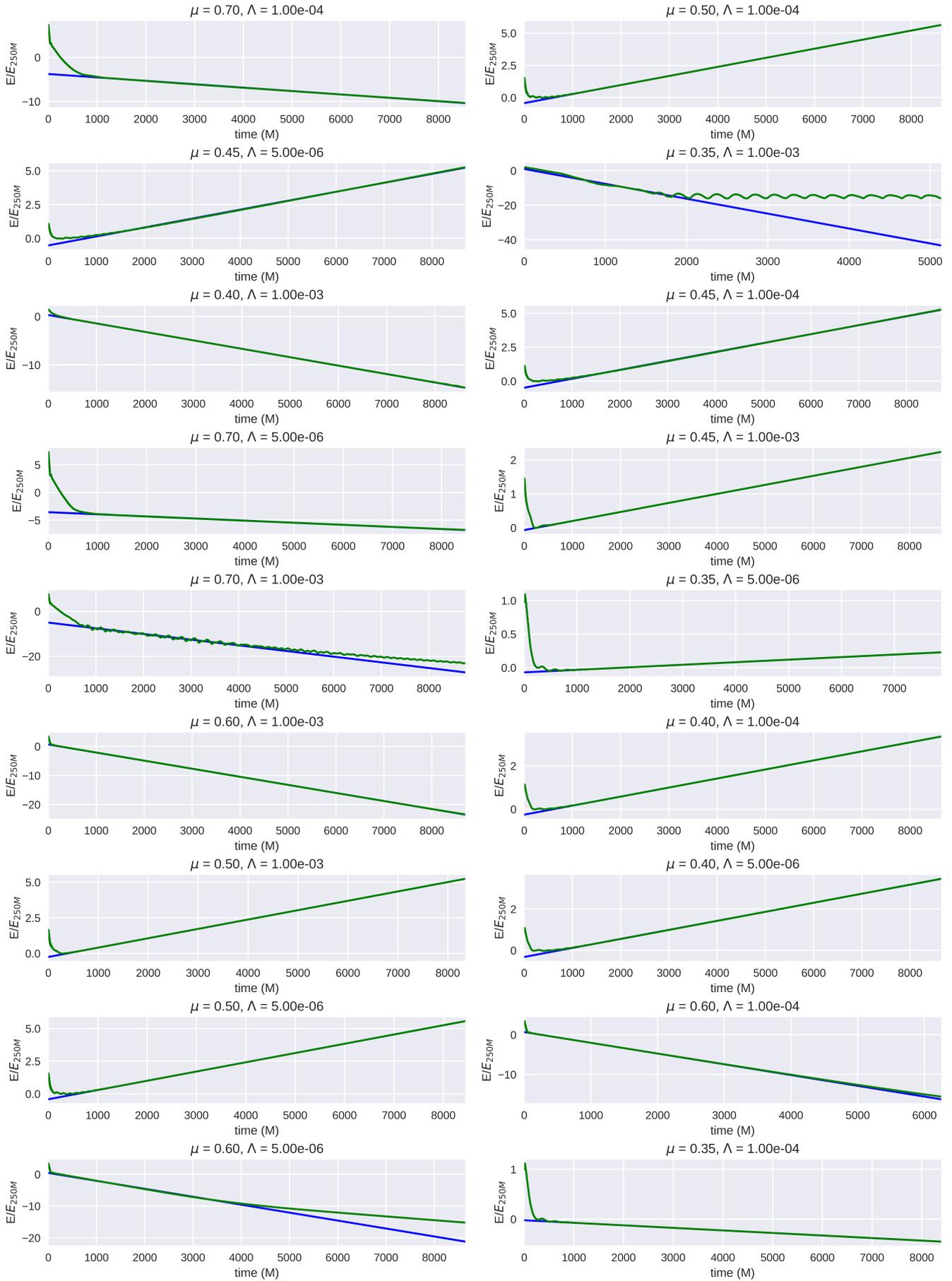}
    \caption{Growth of the energy of the cloud over simulation time. Energy values are normalized to their value at $t=250M$. The energy data is fitted in logarithmic space to a linear function, which captures the exponential characteristics of the instability. The fitting of the data starts at $t=250M$. The green line is the numerical data and the blue straight line is the fit function.  }
    \label{fig:alldata}
\end{figure}

The data from our simulations is available in table~\ref{simdata}. The cosmological constant has been rescaled to $\Lambda = \frac{\bar{\Lambda}}{M^2}$, where $\bar{\Lambda}$ is the unscaled parameter. Plots of the normalized total energy as a function of time are shown in fig.~\ref{fig:alldata}. A curious new feature seems to arise in the case of decaying modes, namely a secondary scale. For example, in the case of $\mu=0.6$ and $\Lambda = 5*10^{-6}$, the decay rate of the total energy slows at around $t=4000M$. Similar features can be seen in the $\mu = 0.7$/$\Lambda=10^{-3}$ plot as well as $\mu=0.6$/$\Lambda=10^{-4}$. Whether this is a numerical artifact or a real emerging scale is uncertain. Analytic studies will need to be performed to determine the true nature of this emerging scale. Due to the variations in the $\mu=0.35$/$\Lambda=10^{-3}$, such a feature cannot be determined. It should be noted that the rapid oscillations and cessation of decay in the $\mu=0.35$/$\Lambda=10^{-3}$ plot is likely purely numerical errors. The energy of the Proca cloud becomes incredibly small, it likely runs in to a precision floor and the simulation becomes numerically meaningless past $t \sim 2000M$.

\begin{table}[ht]
\arrayrulecolor[HTML]{DB5800}
\centering
\begin{tabular}{ |c|p{1cm}|p{3cm}|  }
\hline
\rowcolor{LightCyan}\multicolumn{3}{|c|}{Simulation Data} \\
\hline
$\mu$ & $\Lambda$ & $\omega_i$ \\
\hline
0.45 & 0.000005 & 0.0006612210382883 \\
0.70 & 0.000005 & -0.0003803920672232 \\
0.40 & 0.000005 & 0.0004368389589392 \\
0.35 & 0.000005 & 0.0000378479679286 \\
0.50 & 0.000005 & 0.0007077262886612 \\
0.60 & 0.000005 & -0.0025030874787604 \\
0.50 & 0.0001 & 0.0007037383784603 \\
0.45 & 0.0001 & 0.0006586744541765 \\
0.60 & 0.0001 & -0.0027137433850122 \\
0.40 & 0.0001 & 0.0004173920772338 \\
0.35 & 0.0001 & -0.0000509955398002 \\
0.70 & 0.0001 & -0.0007733224329400 \\
0.45 & 0.001 & 0.0002659457748499 \\
0.70 & 0.001 & -0.0025289981313056 \\
0.35 & 0.001 & -0.0086145055245226 \\
0.40 & 0.001 & -0.0017417580794997 \\
0.50 & 0.001 & 0.0006576997786869 \\
0.60 & 0.001 & -0.0027797013519735 \\
\hline
\end{tabular}
\caption{All available simulation data.}
\label{simdata}
\end{table}

\subsection{Convergence tests} \label{app:convertest}
To ensure the reliability and accuracy of our simulations, we conduct two types of convergence tests. These tests are crucial for validating the fidelity of the simulated data and confirming that it accurately reflects the underlying physical system. The first type of test we perform is the grid resolution test. By running our simulations at multiple grid resolutions, we can verify that the results converge to a stable solution as the grid is refined. This process helps us identify and minimize numerical artifacts that might arise from discretization errors. Specifically, we monitor key physical quantities and ensure that their values stabilize with increasing grid resolution, indicating that our results are not dependent on the grid size. The second type of test is the time step refinement test. Here, we run our simulations with progressively smaller time steps to ensure that the temporal evolution of the system is accurately captured. By comparing results obtained with different time steps, we can check for stability and convergence in the time integration scheme. This test is particularly important for dynamic simulations where rapid changes in the system need to be accurately resolved. In order to maintain the Courant–Friedrichs–Lewy condition, the temporal timestep decreases proportionally with the spatial timestep. Thus, the two checks are unified in our resolution tests. Below, we provide detailed insights into the methodologies employed for conducting these two tests and assess the reliability of our simulations based on their outcomes.

\subsubsection{Resolution Test} \label{app:restest}
The first convergence test is a resolution test. This ensures that finite-differencing effects have a diminishing effect on the data as the resolution is turned up. In other words, the resulting data should converge to stable values as the resolution is increased. After some point, increasing the resolution further should have little effect on the data. This ensures that the resolution we have chosen for our simulations is satisfactory enough to produce good data.

\begin{figure}[hbtp]
    \centering
    \includegraphics[width=1\linewidth]{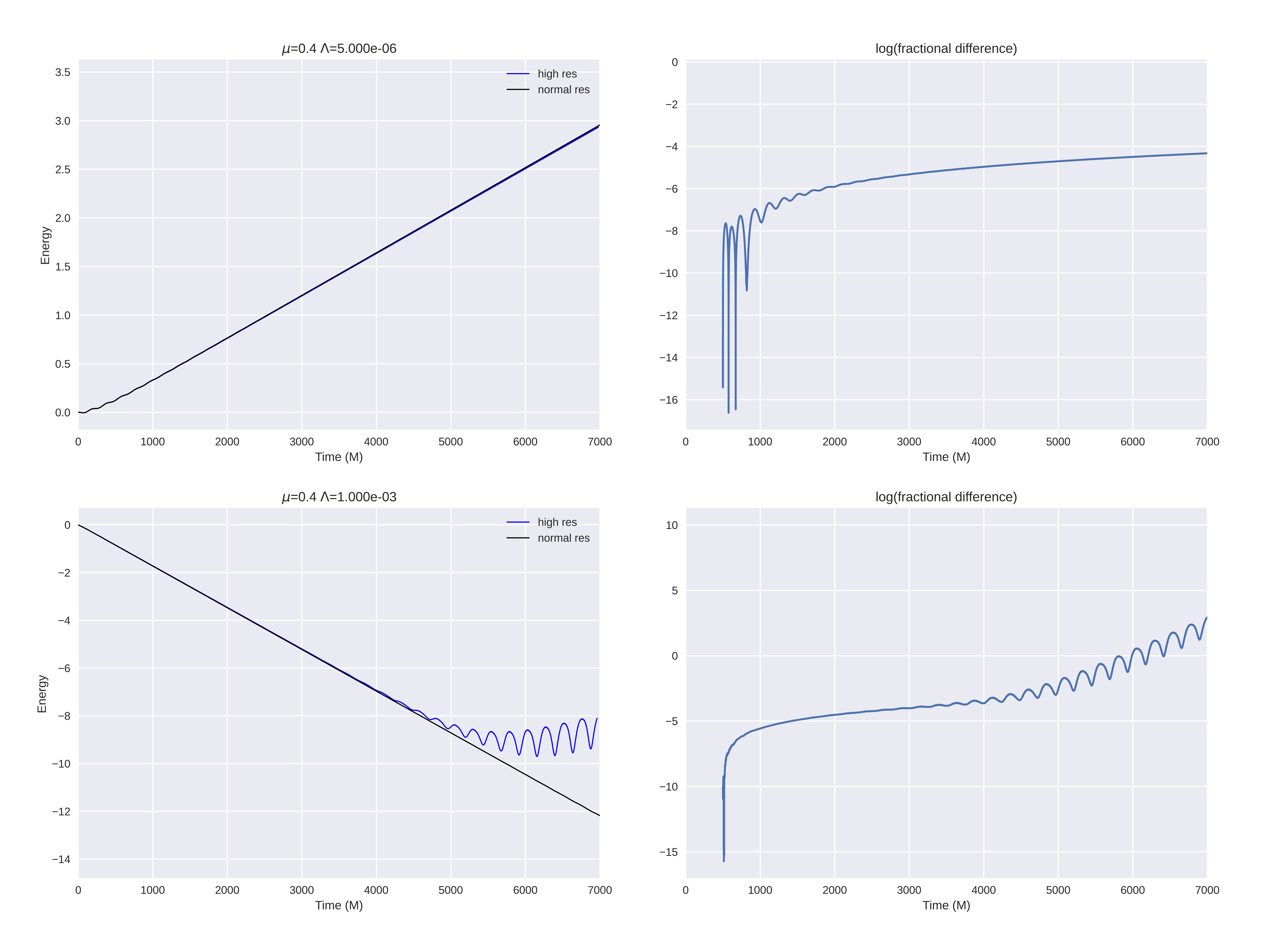}
    \caption{ Test of the convergence of the simulations as resolution is increased. The normal resolution is the resolution used for the main data, which uses $N=192$ grid points. The high resolution simulations use $N=224$ grid points, resulting in a resolution on the finest level of $dx_{fine} = 0.0167M$. The fractional difference is computed as $|E_{norm. res.} - E_{high res.}|/E_{norm. res.}$. The two energy datasets were interpolated over a common time dataset using cubic splines. }
    \label{fig:restest}
\end{figure}

The results of this convergence test are as follow: we repeat two of the simulations with parameters $(\mu,\Lambda) = (0.4, 5*10^{-6})$ and $(0.4, 10^{-3})$ with a resolution on the finest level of $dx_{fine} = 0.0167M$. Since the Courant-Friedrichs-Lewy condition is automatically adjusted for in GRBoondi, the corresponding time resolution is $dt_{fine}=0.0033$, an increase in resolution from the main data, which utilized $dt_{fine} = 0.0039$. It was found that the instability rate for both the growing and decay mode change by less than a percent, signifying good precision for the main data. The two simulations and their differences are shown in fig.~\ref{fig:restest}.

\subsubsection{Analytic Derivatives Test}
The second convergence test is a self-check test on the metric derivatives. Since the procedure in transforming the metric variables and their derivatives from Kerr-Schild coordinates to a Cartesian-like coordinate system is involved, we run a check to make sure the numerically computed derivatives converge to the analytically computed ones from Eq.~\ref{eq:coordtransfup}-\ref{eq:coordtransfdown}.
We follow the procedure of \cite{alcnumrel} for carrying out this convergence test. The test procedure is as follows: 
\begin{center}
	\noindent\resizebox{0.5\textwidth}{!}{
		\begin{tikzpicture}
			[
			main/.style = {draw, rectangle, minimum height=3em, minimum width=10em, fill=blue!20, align=center},
			child/.style = {draw, rectangle, , minimum height=2em, minimum width=2em, fill=green!40, align=center},
			myArrow/.style = {->, arrows={-Stealth[length=6pt,inset=2pt]}, thick, gray},
			myBox/.style={dotted, fill=blue!20, fill opacity=0.3},
			myLine/.style = {dash pattern=on \pgflinewidth off 4pt, line width=.3mm, line cap=round}
			]

			\node[main] (top) { Pick resolution };
			
			\node[child, below=0.9cm of top] (res21) {setup single \\ layer grid};
			\node[child, below=0.1cm of res21] (res22) {set initial data \\ for all variables};
			\node[child, below=0.1cm of res22] (res23) {compute constraints};
			\node[child, below=0.1cm of res23] (res24) {compute numerical\\ metric derivatives};
			\node[child, below=0.1cm of res24] (res25) {compute analytic\\ metric derivatives};
			\node[child, below=0.1cm of res25] (res26) {take differences\\ between numerical\\ and analytic data};
			\node[child, below=0.1cm of res26] (res27) {ensure errors\\ below absolute\\ threshold};
			
			\node[child, left=2cm of res21] (res11) {setup single \\ layer grid};
			\node[child, below=0.1cm of res11] (res12) {set initial data \\ for all variables};
			\node[child, below=0.1cm of res12] (res13) {compute constraints};
			\node[child, below=0.1cm of res13] (res14) {compute numerical\\ metric derivatives};
			\node[child, below=0.1cm of res14] (res15) {compute analytic\\ metric derivatives};
			\node[child, below=0.1cm of res15] (res16)  {take differences\\ between numerical\\ and analytic data};
			\node[child, below=0.1cm of res16] (res17) {ensure errors\\ below absolute\\ threshold};
			
			\node[main, below=of res27] (bottom) {Compute convergence\\ factors $c(t)$};
			
			\coordinate[above=of res11, yshift=-0.8cm] (res1boxtop);
			\coordinate[above=of res21, yshift=-0.8cm] (res2boxtop);
			\path[myArrow] (top) edge["res. 1",swap] (res1boxtop);
			\path[myArrow] (top) edge["res. 2"] (res2boxtop);
			\path[myArrow] (top) edge (3,-1.2);
			\path[myArrow] (top) edge["res. n"] (3.25,-1.2);
			
			\coordinate[below=of res17, yshift=0.7cm] (res1boxbottom);
			\coordinate[below=of res27, yshift=0.7cm] (res2boxbottom);
			\path[myArrow] (res1boxbottom) edge (bottom);
			\path[myArrow] (res2boxbottom) edge (bottom);
			\path[myArrow] (3,-10.2+0.75) edge (bottom);
			\path[myArrow] (3.25,-10.2+0.75) edge (bottom);
			
			
			\coordinate[below=of res24, xshift=2.7cm+0.0cm, yshift=0.75cm](resl);
			\coordinate[below=of res24, xshift=2.7cm + 1.3cm, yshift=0.75cm](resr);
			\path[draw, myLine] (resl) -- (resr);
			

			\begin{scope}[on background layer]
				\coordinate[above left=of res11, xshift=0.2cm,yshift=-0.8cm] (b1l);
				\coordinate[below right=of res17,xshift=-0.5cm, yshift=0.8cm] (b1r);
				\draw[myBox] (b1l) rectangle (b1r);
				
				\coordinate[above left=of res21, xshift=0.2cm,yshift=-0.8cm] (b2l);
				\coordinate[below right=of res27,xshift=-0.5cm, yshift=0.8cm] (b2r);
				\draw[myBox] (b2l) rectangle (b2r);
				
			\end{scope}

		\end{tikzpicture} 
	}
\end{center}

where the convergence factor $c(t)$ is defined as
\begin{equation}
c_i(t) \equiv \frac{ || \epsilon_{\Delta_i} ||}{ || \epsilon_{\Delta_{i+1}} ||}
\end{equation}
and $\Delta_i$ denotes the i'th resolution, $\epsilon_{\Delta_i} = U(t,x) - U_{\Delta_i}(t,x)$, $U(t,x)$ denotes the variable computed using the exact analytic derivatives, and $U_{\Delta_i}(t,x)$ denotes the variable computed using the numerically computed derivatives at a resolution of $\Delta_i$. The resolutions are typically chosen to be twice the previous resolution. In other words, $\Delta_i = 2^i \Delta_1$. This implies that for an nth-order finite differencing scheme, 
\begin{equation}
c(t) = 2^n
\end{equation}
Since GRBoondi uses fourth-order finite differencing stencils for the spatial derivatives, we expect $c(t) = 16$ for each variable. 
We perform the convergence test for two different resolutions, which differ by a multiple of two per the previous discussion. We find that the minimum convergence factor across all grid variables was $14.35$, in fairly good agreement with the expected convergence factor of $16$ for a fourth-order finite differencing routine.

\section{Conclusion}\label{sec:conclusion}
This study investigates the dynamics of a Proca field surrounding a spinning black hole within an expanding universe, a novel approach that advances our understanding of superradiant vector fields interacting with cosmological constants—a crucial yet overlooked aspect in prior research. Leveraging the advanced capabilities of the recently developed GRBoondi software, we conducted a series of simulations aimed at quantifying the growth rates under diverse scenarios involving different Proca masses and cosmological constants. Furthermore, we conducted a rigorous resolution analysis to validate the accuracy and reliability of our simulation data, revealing robust agreement with expected theoretical predictions.

There are a few limitations of the current study. Firstly, the effects of backreaction were ignored, however this approximation is suitable for this analysis as the energy densities were small and gravitational radiation was not desired. Should gravitational emission from the resulting Proca cloud be desired, full numerical computations will need to be performed. Due to GRBoondi outputting checkpoint files in the same format as GRChombo, using GRChombo to perform full computations with initial data from the simulations here will be straightforward. Additionally, many parts of the parameter space were left unsampled, due to the numerical complexities. Primarily, simulating larger cosmological constants is difficult due to the cosmological horizon rapidly approaching the outer horizon of the black hole. Additionally, simulating lower spins increases the simulation time considerably, hence we only focused on a single value of the black hole spin, $\chi=0.99$. A more complete analysis would likely require a different coordinate choice that penetrates the cosmological horizon.

Future studies to be performed include theoretical analyses of various coordinate systems in the search of ones that can penetrate the outer and cosmological horizons. Additionally, future studies will turn to full evolution of the Einstein+Proca system to compute the emitted gravitational radiation and apply the results to gravitational observatory forecasting. An intriguing progression would involve extending beyond the Proca field's mass term to incorporate derivative self-interactions inherent in Generalized Proca theories. This expansion would enhance our exploration of the field's dynamics, encompassing interactions that go beyond simple mass considerations and delve into the complexities introduced by derivative couplings within these theories.

\appendix

\section{Coordinate Transformations} \label{app:coordtransform}
In transforming metric variables from Kerr-Schild to the Cartesian-like coordinate system Eq.~\ref{eq:cartx}-\ref{eq:cartz}, we follow the standard procedure for coordinate transformations. That is
\begin{align}
\beta^{i}_{cart}(x,y,z) &= \Lambda^i_j \beta^j_{KS} (r,\theta,\phi) \\
\gamma_{ij,cart} &= (\Lambda^{-1})_i^k (\Lambda^{-1})_j^l \gamma_{kl,KS}
\end{align}
where $\Lambda^i_j = \frac{ d X^i }{ d R^j}$, $X = (x,y,z)$, and $R = (r,\theta, \phi)$. The derivatives of the metric variables are then computed straight forwardly using the following rule for a rank-r contravariant tensor

\begin{align} \label{eq:coordtransfup}
\partial_p T^{i j \cdots} = \left( \Lambda^{-1} \right)^r_p \left[ \tilde{\partial}_r \left( \Lambda^i_m \Lambda^j_n \cdots \right) \tilde{T}^{m n \cdots} + \left( \Lambda^i_m \Lambda^j_n \cdots \right)  \tilde{\partial}_r \tilde{T}^{m n \cdots} \right]
\end{align}
where a tilde repesents the quantity in the base coordinate system and non-tilde in the new coordinate system. For a covariant rank-r tensor, we find a similar rule
\begin{equation} \label{eq:coordtransfdown}
\partial_r T_{ij \cdots} =  \left( \Lambda^{-1} \right)^p_{r} \left( \left(\Lambda^{-1}\right)^m_i \left(\Lambda^{-1}\right)^n_j \cdots \right) \tilde{\partial}_{p} \tilde{T}_{m n \cdots} + \tilde{T}_{m n \cdots} \partial_r \left( \left(\Lambda^{-1}\right)^m_i \left(\Lambda^{-1}\right)^n_j \cdots \right)
\end{equation}
For example, the spatial metric in the new Cartesian-like coordinate system can be computed as
\begin{equation}
\partial_r \gamma_{ij} = \frac{ \partial \tilde{\gamma}_{kl}}{\partial R^m} \frac{\partial R^m}{\partial X^r} \frac{\partial R^k}{\partial X^i}\frac{\partial R^l}{\partial X^j} + \tilde{\gamma}_{kl} \frac{\partial}{\partial X^r} \left( \frac{\partial R^k}{\partial X^i}\frac{\partial R^l}{\partial X^j} \right)
\end{equation}
The rest of the derivatives of the metric variables follow similarly.
The Jacobian matrix of the transformation Eq.~\ref{eq:cartx}-\ref{eq:cartz} is simple to compute. Since at each grid point, we are given the $(x,y,z)$ coordinates, we represent the Jacobian matrix in terms of the $(x,y,z)$ variables. Hence, it is given by 
\begin{align}
\Lambda^i_j &= 
    \begin{bmatrix}
        \frac{x}{r} & \frac{y}{r} & \frac{z}{r} \\
        \frac{xz}{r^2 \rho} & \frac{yz}{r^2 \rho} & \frac{-\rho}{r^2} \\
        \frac{-y}{\rho^2} & \frac{x}{\rho^2} & 0 
    \end{bmatrix}
\end{align}
where $r^2 = x^2 + y^2 + z^2$ and $\rho^2 = x^2 + y^2$. Due to the division by the two radii, GRBoondi sets a minimum value for the radii of $10^{-12}$. For more details on the numerical implementation, see \cite{Fell_GRBoondi}.

 \begin{acknowledgements}
    The simulations performed in this study utilized the resources of the Baden-W{\"u}rttemberg High Performance Computing cluster. The authors acknowledge support by the state of Baden-Württemberg through bwHPC. L.H. is supported by funding from the European Research Council (ERC) under the European Unions Horizon 2020 research and innovation programme grant agreement No 801781. L.H. further acknowledges support from the Deutsche Forschungsgemeinschaft (DFG, German Research Foundation) under Germany’s Excellence Strategy EXC 2181/1 -390900948 (the Heidelberg STRUCTURES Excellence Cluster).
\end{acknowledgements}

\bibliographystyle{apsrev4-1}
	\bibliography{paper.bib}

\begin{thebibliography}{61}%
\makeatletter
\providecommand \@ifxundefined [1]{%
 \@ifx{#1\undefined}
}%
\providecommand \@ifnum [1]{%
 \ifnum #1\expandafter \@firstoftwo
 \else \expandafter \@secondoftwo
 \fi
}%
\providecommand \@ifx [1]{%
 \ifx #1\expandafter \@firstoftwo
 \else \expandafter \@secondoftwo
 \fi
}%
\providecommand \natexlab [1]{#1}%
\providecommand \enquote  [1]{``#1''}%
\providecommand \bibnamefont  [1]{#1}%
\providecommand \bibfnamefont [1]{#1}%
\providecommand \citenamefont [1]{#1}%
\providecommand \href@noop [0]{\@secondoftwo}%
\providecommand \href [0]{\begingroup \@sanitize@url \@href}%
\providecommand \@href[1]{\@@startlink{#1}\@@href}%
\providecommand \@@href[1]{\endgroup#1\@@endlink}%
\providecommand \@sanitize@url [0]{\catcode `\\12\catcode `\$12\catcode `\&12\catcode `\#12\catcode `\^12\catcode `\_12\catcode `\%12\relax}%
\providecommand \@@startlink[1]{}%
\providecommand \@@endlink[0]{}%
\providecommand \url  [0]{\begingroup\@sanitize@url \@url }%
\providecommand \@url [1]{\endgroup\@href {#1}{\urlprefix }}%
\providecommand \urlprefix  [0]{URL }%
\providecommand \Eprint [0]{\href }%
\providecommand \doibase [0]{http://dx.doi.org/}%
\providecommand \selectlanguage [0]{\@gobble}%
\providecommand \bibinfo  [0]{\@secondoftwo}%
\providecommand \bibfield  [0]{\@secondoftwo}%
\providecommand \translation [1]{[#1]}%
\providecommand \BibitemOpen [0]{}%
\providecommand \bibitemStop [0]{}%
\providecommand \bibitemNoStop [0]{.\EOS\space}%
\providecommand \EOS [0]{\spacefactor3000\relax}%
\providecommand \BibitemShut  [1]{\csname bibitem#1\endcsname}%
\let\auto@bib@innerbib\@empty
\bibitem [{\citenamefont {Heisenberg}(2019)}]{Heisenberg:2018vsk}%
  \BibitemOpen
  \bibfield  {author} {\bibinfo {author} {\bibfnamefont {L.}~\bibnamefont {Heisenberg}},\ }\href {\doibase 10.1016/j.physrep.2018.11.006} {\bibfield  {journal} {\bibinfo  {journal} {Phys. Rept.}\ }\textbf {\bibinfo {volume} {796}},\ \bibinfo {pages} {1} (\bibinfo {year} {2019})},\ \bibinfo {note} {arXiv: 1807.01725},\ \Eprint {http://arxiv.org/abs/1807.01725} {arXiv:1807.01725 [gr-qc]} \BibitemShut {NoStop}%
\bibitem [{\citenamefont {Aghanim}\ \emph {et~al.}(2020)\citenamefont {Aghanim}, \citenamefont {Akrami}, \citenamefont {Ashdown} \emph {et~al.}}]{planck2020}%
  \BibitemOpen
  \bibfield  {author} {\bibinfo {author} {\bibfnamefont {N.}~\bibnamefont {Aghanim}}, \bibinfo {author} {\bibfnamefont {Y.}~\bibnamefont {Akrami}}, \bibinfo {author} {\bibfnamefont {M.}~\bibnamefont {Ashdown}},  \emph {et~al.},\ }\href {\doibase 10.1051/0004-6361/201833910} {\bibfield  {journal} {\bibinfo  {journal} {Astronomy \& Astrophysics}\ }\textbf {\bibinfo {volume} {641}},\ \bibinfo {pages} {A6} (\bibinfo {year} {2020})}\BibitemShut {NoStop}%
\bibitem [{\citenamefont {Riess}\ \emph {et~al.}(2023)\citenamefont {Riess}, \citenamefont {Anand}, \citenamefont {Yuan} \emph {et~al.}}]{riess2023crowded}%
  \BibitemOpen
  \bibfield  {author} {\bibinfo {author} {\bibfnamefont {A.~G.}\ \bibnamefont {Riess}}, \bibinfo {author} {\bibfnamefont {G.~S.}\ \bibnamefont {Anand}}, \bibinfo {author} {\bibfnamefont {W.}~\bibnamefont {Yuan}},  \emph {et~al.},\ }\href@noop {} {\enquote {\bibinfo {title} {Crowded no more: The accuracy of the hubble constant tested with high resolution observations of cepheids by jwst},}\ } (\bibinfo {year} {2023}),\ \Eprint {http://arxiv.org/abs/2307.15806} {arXiv:2307.15806 [astro-ph.CO]} \BibitemShut {NoStop}%
\bibitem [{\citenamefont {Ghirardini}\ \emph {et~al.}(2024)\citenamefont {Ghirardini}, \citenamefont {Bulbul}, \citenamefont {Artis} \emph {et~al.}}]{ghirardini2024srgerosita}%
  \BibitemOpen
  \bibfield  {author} {\bibinfo {author} {\bibfnamefont {V.}~\bibnamefont {Ghirardini}}, \bibinfo {author} {\bibfnamefont {E.}~\bibnamefont {Bulbul}}, \bibinfo {author} {\bibfnamefont {E.}~\bibnamefont {Artis}},  \emph {et~al.},\ }\href@noop {} {\enquote {\bibinfo {title} {The srg/erosita all-sky survey: Cosmology constraints from cluster abundances in the western galactic hemisphere},}\ } (\bibinfo {year} {2024}),\ \Eprint {http://arxiv.org/abs/2402.08458} {arXiv:2402.08458 [astro-ph.CO]} \BibitemShut {NoStop}%
\bibitem [{\citenamefont {{Zwicky}}(2009)}]{2009GReGr..41..207Z}%
  \BibitemOpen
  \bibfield  {author} {\bibinfo {author} {\bibfnamefont {F.}~\bibnamefont {{Zwicky}}},\ }\href {\doibase 10.1007/s10714-008-0707-4} {\bibfield  {journal} {\bibinfo  {journal} {General Relativity and Gravitation}\ }\textbf {\bibinfo {volume} {41}},\ \bibinfo {pages} {207} (\bibinfo {year} {2009})}\BibitemShut {NoStop}%
\bibitem [{\citenamefont {{Zwicky}}(1937)}]{1937ApJ....86..217Z}%
  \BibitemOpen
  \bibfield  {author} {\bibinfo {author} {\bibfnamefont {F.}~\bibnamefont {{Zwicky}}},\ }\href {\doibase 10.1086/143864} {\bibfield  {journal} {\bibinfo  {journal} {\apj}\ }\textbf {\bibinfo {volume} {86}},\ \bibinfo {pages} {217} (\bibinfo {year} {1937})}\BibitemShut {NoStop}%
\bibitem [{\citenamefont {{Smith}}(1936)}]{1936ApJ....83...23S}%
  \BibitemOpen
  \bibfield  {author} {\bibinfo {author} {\bibfnamefont {S.}~\bibnamefont {{Smith}}},\ }\href {\doibase 10.1086/143697} {\bibfield  {journal} {\bibinfo  {journal} {\apj}\ }\textbf {\bibinfo {volume} {83}},\ \bibinfo {pages} {23} (\bibinfo {year} {1936})}\BibitemShut {NoStop}%
\bibitem [{\citenamefont {{Babcock}}(1939)}]{1939LicOB..19...41B}%
  \BibitemOpen
  \bibfield  {author} {\bibinfo {author} {\bibfnamefont {H.~W.}\ \bibnamefont {{Babcock}}},\ }\href {\doibase 10.5479/ADS/bib/1939LicOB.19.41B} {\bibfield  {journal} {\bibinfo  {journal} {Lick Observatory Bulletin}\ }\textbf {\bibinfo {volume} {498}},\ \bibinfo {pages} {41} (\bibinfo {year} {1939})}\BibitemShut {NoStop}%
\bibitem [{\citenamefont {{Rubin}}\ and\ \citenamefont {{Ford}}(1970)}]{1970ApJ...159..379R}%
  \BibitemOpen
  \bibfield  {author} {\bibinfo {author} {\bibfnamefont {V.~C.}\ \bibnamefont {{Rubin}}}\ and\ \bibinfo {author} {\bibfnamefont {J.}~\bibnamefont {{Ford}}, \bibfnamefont {W.~Kent}},\ }\href {\doibase 10.1086/150317} {\bibfield  {journal} {\bibinfo  {journal} {\apj}\ }\textbf {\bibinfo {volume} {159}},\ \bibinfo {pages} {379} (\bibinfo {year} {1970})}\BibitemShut {NoStop}%
\bibitem [{\citenamefont {{Roberts}}\ and\ \citenamefont {{Whitehurst}}(1975)}]{1975ApJ...201..327R}%
  \BibitemOpen
  \bibfield  {author} {\bibinfo {author} {\bibfnamefont {M.~S.}\ \bibnamefont {{Roberts}}}\ and\ \bibinfo {author} {\bibfnamefont {R.~N.}\ \bibnamefont {{Whitehurst}}},\ }\href {\doibase 10.1086/153889} {\bibfield  {journal} {\bibinfo  {journal} {\apj}\ }\textbf {\bibinfo {volume} {201}},\ \bibinfo {pages} {327} (\bibinfo {year} {1975})}\BibitemShut {NoStop}%
\bibitem [{\citenamefont {{Oort}}(1940)}]{1940ApJ....91..273O}%
  \BibitemOpen
  \bibfield  {author} {\bibinfo {author} {\bibfnamefont {J.~H.}\ \bibnamefont {{Oort}}},\ }\href {\doibase 10.1086/144167} {\bibfield  {journal} {\bibinfo  {journal} {\apj}\ }\textbf {\bibinfo {volume} {91}},\ \bibinfo {pages} {273} (\bibinfo {year} {1940})}\BibitemShut {NoStop}%
\bibitem [{\citenamefont {{Rubin}}\ \emph {et~al.}(1985)\citenamefont {{Rubin}}, \citenamefont {{Burstein}}, \citenamefont {{Ford}} \emph {et~al.}}]{1985ApJ...289...81R}%
  \BibitemOpen
  \bibfield  {author} {\bibinfo {author} {\bibfnamefont {V.~C.}\ \bibnamefont {{Rubin}}}, \bibinfo {author} {\bibfnamefont {D.}~\bibnamefont {{Burstein}}}, \bibinfo {author} {\bibfnamefont {J.}~\bibnamefont {{Ford}}, \bibfnamefont {W.~K.}},  \emph {et~al.},\ }\href {\doibase 10.1086/162866} {\bibfield  {journal} {\bibinfo  {journal} {\apj}\ }\textbf {\bibinfo {volume} {289}},\ \bibinfo {pages} {81} (\bibinfo {year} {1985})}\BibitemShut {NoStop}%
\bibitem [{\citenamefont {{Persic}}\ and\ \citenamefont {{Salucci}}(1988)}]{1988MNRAS.234..131P}%
  \BibitemOpen
  \bibfield  {author} {\bibinfo {author} {\bibfnamefont {M.}~\bibnamefont {{Persic}}}\ and\ \bibinfo {author} {\bibfnamefont {P.}~\bibnamefont {{Salucci}}},\ }\href {\doibase 10.1093/mnras/234.1.131} {\bibfield  {journal} {\bibinfo  {journal} {mnras}\ }\textbf {\bibinfo {volume} {234}},\ \bibinfo {pages} {131} (\bibinfo {year} {1988})}\BibitemShut {NoStop}%
\bibitem [{\citenamefont {Einstein}(1916)}]{einsteinrelativity}%
  \BibitemOpen
  \bibfield  {author} {\bibinfo {author} {\bibfnamefont {A.}~\bibnamefont {Einstein}},\ }\href {\doibase https://doi.org/10.1002/andp.19163540702} {\bibfield  {journal} {\bibinfo  {journal} {Annalen der Physik}\ }\textbf {\bibinfo {volume} {354}},\ \bibinfo {pages} {769} (\bibinfo {year} {1916})},\ \Eprint {http://arxiv.org/abs/https://onlinelibrary.wiley.com/doi/pdf/10.1002/andp.19163540702} {https://onlinelibrary.wiley.com/doi/pdf/10.1002/andp.19163540702} \BibitemShut {NoStop}%
\bibitem [{\citenamefont {Aprile}\ \emph {et~al.}(2023)\citenamefont {Aprile}, \citenamefont {Abe}, \citenamefont {Agostini} \emph {et~al.}}]{Aprile_2023}%
  \BibitemOpen
  \bibfield  {author} {\bibinfo {author} {\bibfnamefont {E.}~\bibnamefont {Aprile}}, \bibinfo {author} {\bibfnamefont {K.}~\bibnamefont {Abe}}, \bibinfo {author} {\bibfnamefont {F.}~\bibnamefont {Agostini}},  \emph {et~al.},\ }\href {\doibase 10.1103/physrevlett.131.041003} {\bibfield  {journal} {\bibinfo  {journal} {Physical Review Letters}\ }\textbf {\bibinfo {volume} {131}} (\bibinfo {year} {2023}),\ 10.1103/physrevlett.131.041003}\BibitemShut {NoStop}%
\bibitem [{\citenamefont {Bartram}\ \emph {et~al.}(2021)\citenamefont {Bartram}, \citenamefont {Braine}, \citenamefont {Burns} \emph {et~al.}}]{Bartram_2021}%
  \BibitemOpen
  \bibfield  {author} {\bibinfo {author} {\bibfnamefont {C.}~\bibnamefont {Bartram}}, \bibinfo {author} {\bibfnamefont {T.}~\bibnamefont {Braine}}, \bibinfo {author} {\bibfnamefont {E.}~\bibnamefont {Burns}},  \emph {et~al.},\ }\href {\doibase 10.1103/physrevlett.127.261803} {\bibfield  {journal} {\bibinfo  {journal} {Physical Review Letters}\ }\textbf {\bibinfo {volume} {127}} (\bibinfo {year} {2021}),\ 10.1103/physrevlett.127.261803}\BibitemShut {NoStop}%
\bibitem [{\citenamefont {Jungman}\ \emph {et~al.}(1996)\citenamefont {Jungman}, \citenamefont {Kamionkowski},\ and\ \citenamefont {Griest}}]{Jungman1996}%
  \BibitemOpen
  \bibfield  {author} {\bibinfo {author} {\bibfnamefont {G.}~\bibnamefont {Jungman}}, \bibinfo {author} {\bibfnamefont {M.}~\bibnamefont {Kamionkowski}}, \ and\ \bibinfo {author} {\bibfnamefont {K.}~\bibnamefont {Griest}},\ }\href {\doibase 10.1016/0370-1573(95)00058-5} {\bibfield  {journal} {\bibinfo  {journal} {Physics Reports}\ }\textbf {\bibinfo {volume} {267}},\ \bibinfo {pages} {195–373} (\bibinfo {year} {1996})}\BibitemShut {NoStop}%
\bibitem [{\citenamefont {Preskill}\ \emph {et~al.}(1983)\citenamefont {Preskill}, \citenamefont {Wise},\ and\ \citenamefont {Wilczek}}]{Preskill1983}%
  \BibitemOpen
  \bibfield  {author} {\bibinfo {author} {\bibfnamefont {J.}~\bibnamefont {Preskill}}, \bibinfo {author} {\bibfnamefont {M.~B.}\ \bibnamefont {Wise}}, \ and\ \bibinfo {author} {\bibfnamefont {F.}~\bibnamefont {Wilczek}},\ }\href {\doibase 10.1016/0370-2693(83)90637-8} {\bibfield  {journal} {\bibinfo  {journal} {Physics Letters B}\ }\textbf {\bibinfo {volume} {120}},\ \bibinfo {pages} {127–132} (\bibinfo {year} {1983})}\BibitemShut {NoStop}%
\bibitem [{\citenamefont {Essig}\ \emph {et~al.}(2013)\citenamefont {Essig}, \citenamefont {Jaros}, \citenamefont {Wester} \emph {et~al.}}]{essig2013dark}%
  \BibitemOpen
  \bibfield  {author} {\bibinfo {author} {\bibfnamefont {R.}~\bibnamefont {Essig}}, \bibinfo {author} {\bibfnamefont {J.~A.}\ \bibnamefont {Jaros}}, \bibinfo {author} {\bibfnamefont {W.}~\bibnamefont {Wester}},  \emph {et~al.},\ }\href@noop {} {\enquote {\bibinfo {title} {Dark sectors and new, light, weakly-coupled particles},}\ } (\bibinfo {year} {2013}),\ \Eprint {http://arxiv.org/abs/1311.0029} {arXiv:1311.0029 [hep-ph]} \BibitemShut {NoStop}%
\bibitem [{\citenamefont {Holdom}(1986)}]{Holdom1986}%
  \BibitemOpen
  \bibfield  {author} {\bibinfo {author} {\bibfnamefont {B.}~\bibnamefont {Holdom}},\ }\href {\doibase 10.1016/0370-2693(86)91377-8} {\bibfield  {journal} {\bibinfo  {journal} {Physics Letters B}\ }\textbf {\bibinfo {volume} {166}},\ \bibinfo {pages} {196–198} (\bibinfo {year} {1986})}\BibitemShut {NoStop}%
\bibitem [{\citenamefont {Galison}\ and\ \citenamefont {Manohar}(1984)}]{Galison1984}%
  \BibitemOpen
  \bibfield  {author} {\bibinfo {author} {\bibfnamefont {P.}~\bibnamefont {Galison}}\ and\ \bibinfo {author} {\bibfnamefont {A.}~\bibnamefont {Manohar}},\ }\href {\doibase 10.1016/0370-2693(84)91161-4} {\bibfield  {journal} {\bibinfo  {journal} {Physics Letters B}\ }\textbf {\bibinfo {volume} {136}},\ \bibinfo {pages} {279–283} (\bibinfo {year} {1984})}\BibitemShut {NoStop}%
\bibitem [{\citenamefont {Nelson}\ and\ \citenamefont {Scholtz}(2011)}]{Nelson_Scholtz_2011}%
  \BibitemOpen
  \bibfield  {author} {\bibinfo {author} {\bibfnamefont {A.~E.}\ \bibnamefont {Nelson}}\ and\ \bibinfo {author} {\bibfnamefont {J.}~\bibnamefont {Scholtz}},\ }\href {\doibase 10.1103/PhysRevD.84.103501} {\bibfield  {journal} {\bibinfo  {journal} {Physical Review D}\ }\textbf {\bibinfo {volume} {84}},\ \bibinfo {pages} {103501} (\bibinfo {year} {2011})},\ \bibinfo {note} {arXiv:1105.2812 [astro-ph, physics:hep-ph]}\BibitemShut {NoStop}%
\bibitem [{\citenamefont {Goodsell}\ \emph {et~al.}(2009)\citenamefont {Goodsell}, \citenamefont {Jaeckel}, \citenamefont {Redondo} \emph {et~al.}}]{Goodsell_Jaeckel_Redondo_Ringwald_2009}%
  \BibitemOpen
  \bibfield  {author} {\bibinfo {author} {\bibfnamefont {M.}~\bibnamefont {Goodsell}}, \bibinfo {author} {\bibfnamefont {J.}~\bibnamefont {Jaeckel}}, \bibinfo {author} {\bibfnamefont {J.}~\bibnamefont {Redondo}},  \emph {et~al.},\ }\href {\doibase 10.1088/1126-6708/2009/11/027} {\bibfield  {journal} {\bibinfo  {journal} {Journal of High Energy Physics}\ }\textbf {\bibinfo {volume} {2009}},\ \bibinfo {pages} {027–027} (\bibinfo {year} {2009})},\ \bibinfo {note} {arXiv:0909.0515 [hep-ph, physics:hep-th]}\BibitemShut {NoStop}%
\bibitem [{\citenamefont {Co}\ \emph {et~al.}(2019)\citenamefont {Co}, \citenamefont {Pierce}, \citenamefont {Zhang} \emph {et~al.}}]{Co_2019}%
  \BibitemOpen
  \bibfield  {author} {\bibinfo {author} {\bibfnamefont {R.~T.}\ \bibnamefont {Co}}, \bibinfo {author} {\bibfnamefont {A.}~\bibnamefont {Pierce}}, \bibinfo {author} {\bibfnamefont {Z.}~\bibnamefont {Zhang}},  \emph {et~al.},\ }\href {\doibase 10.1103/physrevd.99.075002} {\bibfield  {journal} {\bibinfo  {journal} {Physical Review D}\ }\textbf {\bibinfo {volume} {99}} (\bibinfo {year} {2019}),\ 10.1103/physrevd.99.075002}\BibitemShut {NoStop}%
\bibitem [{\citenamefont {Graham}\ \emph {et~al.}(2016)\citenamefont {Graham}, \citenamefont {Mardon},\ and\ \citenamefont {Rajendran}}]{Graham_2016}%
  \BibitemOpen
  \bibfield  {author} {\bibinfo {author} {\bibfnamefont {P.~W.}\ \bibnamefont {Graham}}, \bibinfo {author} {\bibfnamefont {J.}~\bibnamefont {Mardon}}, \ and\ \bibinfo {author} {\bibfnamefont {S.}~\bibnamefont {Rajendran}},\ }\href {\doibase 10.1103/physrevd.93.103520} {\bibfield  {journal} {\bibinfo  {journal} {Physical Review D}\ }\textbf {\bibinfo {volume} {93}} (\bibinfo {year} {2016}),\ 10.1103/physrevd.93.103520}\BibitemShut {NoStop}%
\bibitem [{\citenamefont {Su}\ \emph {et~al.}(1994)\citenamefont {Su}, \citenamefont {Heckel}, \citenamefont {Adelberger} \emph {et~al.}}]{PhysRevD.50.3614}%
  \BibitemOpen
  \bibfield  {author} {\bibinfo {author} {\bibfnamefont {Y.}~\bibnamefont {Su}}, \bibinfo {author} {\bibfnamefont {B.~R.}\ \bibnamefont {Heckel}}, \bibinfo {author} {\bibfnamefont {E.~G.}\ \bibnamefont {Adelberger}},  \emph {et~al.},\ }\href {\doibase 10.1103/PhysRevD.50.3614} {\bibfield  {journal} {\bibinfo  {journal} {Phys. Rev. D}\ }\textbf {\bibinfo {volume} {50}},\ \bibinfo {pages} {3614} (\bibinfo {year} {1994})}\BibitemShut {NoStop}%
\bibitem [{\citenamefont {Schlamminger}\ \emph {et~al.}(2008)\citenamefont {Schlamminger}, \citenamefont {Choi}, \citenamefont {Wagner} \emph {et~al.}}]{Schlamminger_2008}%
  \BibitemOpen
  \bibfield  {author} {\bibinfo {author} {\bibfnamefont {S.}~\bibnamefont {Schlamminger}}, \bibinfo {author} {\bibfnamefont {K.-Y.}\ \bibnamefont {Choi}}, \bibinfo {author} {\bibfnamefont {T.~A.}\ \bibnamefont {Wagner}},  \emph {et~al.},\ }\href {\doibase 10.1103/physrevlett.100.041101} {\bibfield  {journal} {\bibinfo  {journal} {Physical Review Letters}\ }\textbf {\bibinfo {volume} {100}} (\bibinfo {year} {2008}),\ 10.1103/physrevlett.100.041101}\BibitemShut {NoStop}%
\bibitem [{\citenamefont {Williams}\ \emph {et~al.}(2004)\citenamefont {Williams}, \citenamefont {Turyshev},\ and\ \citenamefont {Boggs}}]{PhysRevLett.93.261101}%
  \BibitemOpen
  \bibfield  {author} {\bibinfo {author} {\bibfnamefont {J.~G.}\ \bibnamefont {Williams}}, \bibinfo {author} {\bibfnamefont {S.~G.}\ \bibnamefont {Turyshev}}, \ and\ \bibinfo {author} {\bibfnamefont {D.~H.}\ \bibnamefont {Boggs}},\ }\href {\doibase 10.1103/PhysRevLett.93.261101} {\bibfield  {journal} {\bibinfo  {journal} {Phys. Rev. Lett.}\ }\textbf {\bibinfo {volume} {93}},\ \bibinfo {pages} {261101} (\bibinfo {year} {2004})}\BibitemShut {NoStop}%
\bibitem [{\citenamefont {TURYSHEV}\ and\ \citenamefont {WILLIAMS}(2007)}]{TURYSHEV_2007}%
  \BibitemOpen
  \bibfield  {author} {\bibinfo {author} {\bibfnamefont {S.~G.}\ \bibnamefont {TURYSHEV}}\ and\ \bibinfo {author} {\bibfnamefont {J.~G.}\ \bibnamefont {WILLIAMS}},\ }\href {\doibase 10.1142/s0218271807011838} {\bibfield  {journal} {\bibinfo  {journal} {International Journal of Modern Physics D}\ }\textbf {\bibinfo {volume} {16}},\ \bibinfo {pages} {2165} (\bibinfo {year} {2007})}\BibitemShut {NoStop}%
\bibitem [{\citenamefont {Baryakhtar}\ \emph {et~al.}(2017)\citenamefont {Baryakhtar}, \citenamefont {Lasenby},\ and\ \citenamefont {Teo}}]{Baryakhtar_Lasenby_Teo_2017}%
  \BibitemOpen
  \bibfield  {author} {\bibinfo {author} {\bibfnamefont {M.}~\bibnamefont {Baryakhtar}}, \bibinfo {author} {\bibfnamefont {R.}~\bibnamefont {Lasenby}}, \ and\ \bibinfo {author} {\bibfnamefont {M.}~\bibnamefont {Teo}},\ }\href {\doibase 10.1103/PhysRevD.96.035019} {\bibfield  {journal} {\bibinfo  {journal} {Physical Review D}\ }\textbf {\bibinfo {volume} {96}},\ \bibinfo {pages} {035019} (\bibinfo {year} {2017})},\ \bibinfo {note} {arXiv: 1704.05081}\BibitemShut {NoStop}%
\bibitem [{\citenamefont {East}\ and\ \citenamefont {Pretorius}(2017)}]{East_Pretorius_2017}%
  \BibitemOpen
  \bibfield  {author} {\bibinfo {author} {\bibfnamefont {W.~E.}\ \bibnamefont {East}}\ and\ \bibinfo {author} {\bibfnamefont {F.}~\bibnamefont {Pretorius}},\ }\href {\doibase 10.1103/PhysRevLett.119.041101} {\bibfield  {journal} {\bibinfo  {journal} {Physical Review Letters}\ }\textbf {\bibinfo {volume} {119}},\ \bibinfo {pages} {041101} (\bibinfo {year} {2017})}\BibitemShut {NoStop}%
\bibitem [{\citenamefont {East}(2017)}]{East_2017}%
  \BibitemOpen
  \bibfield  {author} {\bibinfo {author} {\bibfnamefont {W.~E.}\ \bibnamefont {East}},\ }\href {\doibase 10.1103/physrevd.96.024004} {\bibfield  {journal} {\bibinfo  {journal} {Physical Review D}\ }\textbf {\bibinfo {volume} {96}},\ \bibinfo {pages} {024004} (\bibinfo {year} {2017})},\ \bibinfo {note} {arXiv:1705.01544 [astro-ph, physics:gr-qc, physics:hep-ph]}\BibitemShut {NoStop}%
\bibitem [{\citenamefont {Pierce}\ \emph {et~al.}(2018)\citenamefont {Pierce}, \citenamefont {Riles},\ and\ \citenamefont {Zhao}}]{Pierce_Riles_Zhao_2018}%
  \BibitemOpen
  \bibfield  {author} {\bibinfo {author} {\bibfnamefont {A.}~\bibnamefont {Pierce}}, \bibinfo {author} {\bibfnamefont {K.}~\bibnamefont {Riles}}, \ and\ \bibinfo {author} {\bibfnamefont {Y.}~\bibnamefont {Zhao}},\ }\href {\doibase 10.1103/PhysRevLett.121.061102} {\bibfield  {journal} {\bibinfo  {journal} {Physical Review Letters}\ }\textbf {\bibinfo {volume} {121}},\ \bibinfo {pages} {061102} (\bibinfo {year} {2018})},\ \bibinfo {note} {arXiv:1801.10161 [astro-ph, physics:gr-qc, physics:hep-ph]}\BibitemShut {NoStop}%
\bibitem [{\citenamefont {Heisenberg}(2014)}]{Heisenberg:2014rta}%
  \BibitemOpen
  \bibfield  {author} {\bibinfo {author} {\bibfnamefont {L.}~\bibnamefont {Heisenberg}},\ }\href {\doibase 10.1088/1475-7516/2014/05/015} {\bibfield  {journal} {\bibinfo  {journal} {JCAP}\ }\textbf {\bibinfo {volume} {05}},\ \bibinfo {pages} {015} (\bibinfo {year} {2014})},\ \bibinfo {note} {arXiv: 1402.7026},\ \Eprint {http://arxiv.org/abs/1402.7026} {arXiv:1402.7026 [hep-th]} \BibitemShut {NoStop}%
\bibitem [{\citenamefont {De~Felice}\ \emph {et~al.}(2016{\natexlab{a}})\citenamefont {De~Felice}, \citenamefont {Heisenberg}, \citenamefont {Kase} \emph {et~al.}}]{DeFelice:2016yws}%
  \BibitemOpen
  \bibfield  {author} {\bibinfo {author} {\bibfnamefont {A.}~\bibnamefont {De~Felice}}, \bibinfo {author} {\bibfnamefont {L.}~\bibnamefont {Heisenberg}}, \bibinfo {author} {\bibfnamefont {R.}~\bibnamefont {Kase}},  \emph {et~al.},\ }\href {\doibase 10.1088/1475-7516/2016/06/048} {\bibfield  {journal} {\bibinfo  {journal} {JCAP}\ }\textbf {\bibinfo {volume} {06}},\ \bibinfo {pages} {048} (\bibinfo {year} {2016}{\natexlab{a}})},\ \Eprint {http://arxiv.org/abs/1603.05806} {arXiv:1603.05806 [gr-qc]} \BibitemShut {NoStop}%
\bibitem [{\citenamefont {De~Felice}\ \emph {et~al.}(2016{\natexlab{b}})\citenamefont {De~Felice}, \citenamefont {Heisenberg}, \citenamefont {Kase} \emph {et~al.}}]{DeFelice:2016uil}%
  \BibitemOpen
  \bibfield  {author} {\bibinfo {author} {\bibfnamefont {A.}~\bibnamefont {De~Felice}}, \bibinfo {author} {\bibfnamefont {L.}~\bibnamefont {Heisenberg}}, \bibinfo {author} {\bibfnamefont {R.}~\bibnamefont {Kase}},  \emph {et~al.},\ }\href {\doibase 10.1103/PhysRevD.94.044024} {\bibfield  {journal} {\bibinfo  {journal} {Phys. Rev. D}\ }\textbf {\bibinfo {volume} {94}},\ \bibinfo {pages} {044024} (\bibinfo {year} {2016}{\natexlab{b}})},\ \Eprint {http://arxiv.org/abs/1605.05066} {arXiv:1605.05066 [gr-qc]} \BibitemShut {NoStop}%
\bibitem [{\citenamefont {Brito}\ \emph {et~al.}(2015)\citenamefont {Brito}, \citenamefont {Cardoso},\ and\ \citenamefont {Pani}}]{Brito_2015}%
  \BibitemOpen
  \bibfield  {author} {\bibinfo {author} {\bibfnamefont {R.}~\bibnamefont {Brito}}, \bibinfo {author} {\bibfnamefont {V.}~\bibnamefont {Cardoso}}, \ and\ \bibinfo {author} {\bibfnamefont {P.}~\bibnamefont {Pani}},\ }\href {\doibase 10.1088/0264-9381/32/13/134001} {\bibfield  {journal} {\bibinfo  {journal} {Classical and Quantum Gravity}\ }\textbf {\bibinfo {volume} {32}},\ \bibinfo {pages} {134001} (\bibinfo {year} {2015})}\BibitemShut {NoStop}%
\bibitem [{\citenamefont {Cardoso}\ \emph {et~al.}(2018)\citenamefont {Cardoso}, \citenamefont {Dias}, \citenamefont {Hartnett} \emph {et~al.}}]{Cardoso_2018}%
  \BibitemOpen
  \bibfield  {author} {\bibinfo {author} {\bibfnamefont {V.}~\bibnamefont {Cardoso}}, \bibinfo {author} {\bibfnamefont {O.~J.}\ \bibnamefont {Dias}}, \bibinfo {author} {\bibfnamefont {G.~S.}\ \bibnamefont {Hartnett}},  \emph {et~al.},\ }\href {\doibase 10.1088/1475-7516/2018/03/043} {\bibfield  {journal} {\bibinfo  {journal} {Journal of Cosmology and Astroparticle Physics}\ }\textbf {\bibinfo {volume} {2018}},\ \bibinfo {pages} {043–043} (\bibinfo {year} {2018})}\BibitemShut {NoStop}%
\bibitem [{\citenamefont {Siemonsen}\ and\ \citenamefont {East}(2020)}]{Siemonsen_East_2020}%
  \BibitemOpen
  \bibfield  {author} {\bibinfo {author} {\bibfnamefont {N.}~\bibnamefont {Siemonsen}}\ and\ \bibinfo {author} {\bibfnamefont {W.~E.}\ \bibnamefont {East}},\ }\href {\doibase 10.1103/PhysRevD.101.024019} {\bibfield  {journal} {\bibinfo  {journal} {Physical Review D}\ }\textbf {\bibinfo {volume} {101}},\ \bibinfo {pages} {024019} (\bibinfo {year} {2020})},\ \bibinfo {note} {arXiv: 1910.09476}\BibitemShut {NoStop}%
\bibitem [{\citenamefont {East}(2018)}]{East_2018}%
  \BibitemOpen
  \bibfield  {author} {\bibinfo {author} {\bibfnamefont {W.~E.}\ \bibnamefont {East}},\ }\href {\doibase 10.1103/physrevlett.121.131104} {\bibfield  {journal} {\bibinfo  {journal} {Physical Review Letters}\ }\textbf {\bibinfo {volume} {121}},\ \bibinfo {pages} {131104} (\bibinfo {year} {2018})},\ \bibinfo {note} {arXiv:1807.00043 [astro-ph, physics:gr-qc, physics:hep-ph]}\BibitemShut {NoStop}%
\bibitem [{\citenamefont {Fell}\ \emph {et~al.}(2023)\citenamefont {Fell}, \citenamefont {Heisenberg},\ and\ \citenamefont {Veske}}]{PhysRevD.108.083010}%
  \BibitemOpen
  \bibfield  {author} {\bibinfo {author} {\bibfnamefont {S.}~\bibnamefont {Fell}}, \bibinfo {author} {\bibfnamefont {L.}~\bibnamefont {Heisenberg}}, \ and\ \bibinfo {author} {\bibfnamefont {D.~b.~u.}\ \bibnamefont {Veske}},\ }\href {\doibase 10.1103/PhysRevD.108.083010} {\bibfield  {journal} {\bibinfo  {journal} {Phys. Rev. D}\ }\textbf {\bibinfo {volume} {108}},\ \bibinfo {pages} {083010} (\bibinfo {year} {2023})}\BibitemShut {NoStop}%
\bibitem [{\citenamefont {Danzmann}(2000)}]{lisa1}%
  \BibitemOpen
  \bibfield  {author} {\bibinfo {author} {\bibfnamefont {K.}~\bibnamefont {Danzmann}},\ }\href {\doibase 10.1016/S0273-1177(99)00973-4} {\bibfield  {journal} {\bibinfo  {journal} {Advances in Space Research}\ }\textbf {\bibinfo {volume} {25}},\ \bibinfo {pages} {1129–1136} (\bibinfo {year} {2000})}\BibitemShut {NoStop}%
\bibitem [{\citenamefont {Amaro-Seoane}\ \emph {et~al.}(2017)\citenamefont {Amaro-Seoane}, \citenamefont {Audley}, \citenamefont {Babak} \emph {et~al.}}]{lisa2}%
  \BibitemOpen
  \bibfield  {author} {\bibinfo {author} {\bibfnamefont {P.}~\bibnamefont {Amaro-Seoane}}, \bibinfo {author} {\bibfnamefont {H.}~\bibnamefont {Audley}}, \bibinfo {author} {\bibfnamefont {S.}~\bibnamefont {Babak}},  \emph {et~al.},\ }\href {\doibase 10.48550/ARXIV.1702.00786} {\enquote {\bibinfo {title} {Laser interferometer space antenna},}\ } (\bibinfo {year} {2017})\BibitemShut {NoStop}%
\bibitem [{\citenamefont {Baumann}\ \emph {et~al.}(2019)\citenamefont {Baumann}, \citenamefont {Chia}, \citenamefont {Stout} \emph {et~al.}}]{Baumann_Chia_Stout_ter}%
  \BibitemOpen
  \bibfield  {author} {\bibinfo {author} {\bibfnamefont {D.}~\bibnamefont {Baumann}}, \bibinfo {author} {\bibfnamefont {H.~S.}\ \bibnamefont {Chia}}, \bibinfo {author} {\bibfnamefont {J.}~\bibnamefont {Stout}},  \emph {et~al.},\ }\href {\doibase 10.1088/1475-7516/2019/12/006} {\bibfield  {journal} {\bibinfo  {journal} {Journal of Cosmology and Astroparticle Physics}\ }\textbf {\bibinfo {volume} {2019}},\ \bibinfo {pages} {006–006} (\bibinfo {year} {2019})},\ \bibinfo {note} {arXiv: 1908.10370}\BibitemShut {NoStop}%
\bibitem [{\citenamefont {Rosa}\ and\ \citenamefont {Dolan}(2012)}]{Rosa_Dolan_2012}%
  \BibitemOpen
  \bibfield  {author} {\bibinfo {author} {\bibfnamefont {J.~G.}\ \bibnamefont {Rosa}}\ and\ \bibinfo {author} {\bibfnamefont {S.~R.}\ \bibnamefont {Dolan}},\ }\href {\doibase 10.1103/PhysRevD.85.044043} {\bibfield  {journal} {\bibinfo  {journal} {Physical Review D}\ }\textbf {\bibinfo {volume} {85}},\ \bibinfo {pages} {044043} (\bibinfo {year} {2012})},\ \bibinfo {note} {arXiv: 1110.4494}\BibitemShut {NoStop}%
\bibitem [{\citenamefont {Frolov}\ \emph {et~al.}(2018)\citenamefont {Frolov}, \citenamefont {Krtou\v{s}}, \citenamefont {Kubiz\v{n}\'{a}k} \emph {et~al.}}]{FKKS_2018}%
  \BibitemOpen
  \bibfield  {author} {\bibinfo {author} {\bibfnamefont {V.~P.}\ \bibnamefont {Frolov}}, \bibinfo {author} {\bibfnamefont {P.}~\bibnamefont {Krtou\v{s}}}, \bibinfo {author} {\bibfnamefont {D.}~\bibnamefont {Kubiz\v{n}\'{a}k}},  \emph {et~al.},\ }\href {\doibase 10.1103/PhysRevLett.120.231103} {\bibfield  {journal} {\bibinfo  {journal} {Physical Review Letters}\ }\textbf {\bibinfo {volume} {120}},\ \bibinfo {pages} {231103} (\bibinfo {year} {2018})},\ \bibinfo {note} {arXiv: 1804.00030}\BibitemShut {NoStop}%
\bibitem [{\citenamefont {Herdeiro}\ \emph {et~al.}(2016)\citenamefont {Herdeiro}, \citenamefont {Radu},\ and\ \citenamefont {Runarsson}}]{Herdeiro_Radu_Runarsson_2016}%
  \BibitemOpen
  \bibfield  {author} {\bibinfo {author} {\bibfnamefont {C.}~\bibnamefont {Herdeiro}}, \bibinfo {author} {\bibfnamefont {E.}~\bibnamefont {Radu}}, \ and\ \bibinfo {author} {\bibfnamefont {H.}~\bibnamefont {Runarsson}},\ }\href {\doibase 10.1088/0264-9381/33/15/154001} {\bibfield  {journal} {\bibinfo  {journal} {Classical and Quantum Gravity}\ }\textbf {\bibinfo {volume} {33}},\ \bibinfo {pages} {154001} (\bibinfo {year} {2016})},\ \bibinfo {note} {arXiv:1603.02687 [astro-ph, physics:gr-qc, physics:hep-th]}\BibitemShut {NoStop}%
\bibitem [{\citenamefont {Dolan}(2018)}]{Dolan_2018}%
  \BibitemOpen
  \bibfield  {author} {\bibinfo {author} {\bibfnamefont {S.~R.}\ \bibnamefont {Dolan}},\ }\href {\doibase 10.1103/PhysRevD.98.104006} {\bibfield  {journal} {\bibinfo  {journal} {Physical Review D}\ }\textbf {\bibinfo {volume} {98}},\ \bibinfo {pages} {104006} (\bibinfo {year} {2018})},\ \bibinfo {note} {arXiv: 1806.01604}\BibitemShut {NoStop}%
\bibitem [{\citenamefont {Cutler}\ \emph {et~al.}(1994)\citenamefont {Cutler}, \citenamefont {Kennefick},\ and\ \citenamefont {Poisson}}]{Cutler_Kennefick_Poisson_1994}%
  \BibitemOpen
  \bibfield  {author} {\bibinfo {author} {\bibfnamefont {C.}~\bibnamefont {Cutler}}, \bibinfo {author} {\bibfnamefont {D.}~\bibnamefont {Kennefick}}, \ and\ \bibinfo {author} {\bibfnamefont {E.}~\bibnamefont {Poisson}},\ }\href {\doibase 10.1103/PhysRevD.50.3816} {\bibfield  {journal} {\bibinfo  {journal} {Physical Review D}\ }\textbf {\bibinfo {volume} {50}},\ \bibinfo {pages} {3816–3835} (\bibinfo {year} {1994})}\BibitemShut {NoStop}%
\bibitem [{\citenamefont {Yoshino}\ and\ \citenamefont {Kodama}(2014)}]{Yoshino_Kodama_2014}%
  \BibitemOpen
  \bibfield  {author} {\bibinfo {author} {\bibfnamefont {H.}~\bibnamefont {Yoshino}}\ and\ \bibinfo {author} {\bibfnamefont {H.}~\bibnamefont {Kodama}},\ }\href {\doibase 10.1093/ptep/ptu029} {\bibfield  {journal} {\bibinfo  {journal} {Progress of Theoretical and Experimental Physics}\ }\textbf {\bibinfo {volume} {2014}} (\bibinfo {year} {2014}),\ 10.1093/ptep/ptu029},\ \bibinfo {note} {arXiv:1312.2326 [gr-qc, physics:hep-ph, physics:hep-th]}\BibitemShut {NoStop}%
\bibitem [{\citenamefont {Santos}\ \emph {et~al.}(2020)\citenamefont {Santos}, \citenamefont {Benone}, \citenamefont {Crispino} \emph {et~al.}}]{Santos_Benone_Crispino_Herdeiro_Radu_2020}%
  \BibitemOpen
  \bibfield  {author} {\bibinfo {author} {\bibfnamefont {N.~M.}\ \bibnamefont {Santos}}, \bibinfo {author} {\bibfnamefont {C.~L.}\ \bibnamefont {Benone}}, \bibinfo {author} {\bibfnamefont {L.~C.~B.}\ \bibnamefont {Crispino}},  \emph {et~al.},\ }\href {\doibase 10.1007/JHEP07(2020)010} {\bibfield  {journal} {\bibinfo  {journal} {Journal of High Energy Physics}\ }\textbf {\bibinfo {volume} {2020}},\ \bibinfo {pages} {10} (\bibinfo {year} {2020})},\ \bibinfo {note} {arXiv: 2004.09536}\BibitemShut {NoStop}%
\bibitem [{\citenamefont {Griffiths}\ and\ \citenamefont {Podolsk\'{y}}(2009)}]{Griffiths_Podolsky_2009}%
  \BibitemOpen
  \bibfield  {author} {\bibinfo {author} {\bibfnamefont {J.~B.}\ \bibnamefont {Griffiths}}\ and\ \bibinfo {author} {\bibfnamefont {J.}~\bibnamefont {Podolsk\'{y}}},\ }\enquote {\bibinfo {title} {Pleba\'{n}ski–demia\'{n}ski solutions},}\ in\ \href@noop {} {\emph {\bibinfo {booktitle} {Exact Space-Times in Einstein's General Relativity}}},\ \bibinfo {series and number} {Cambridge Monographs on Mathematical Physics}\ (\bibinfo  {publisher} {Cambridge University Press},\ \bibinfo {year} {2009})\ p.\ \bibinfo {pages} {304–322}\BibitemShut {NoStop}%
\bibitem [{\citenamefont {Plebanski}\ and\ \citenamefont {Demianski}(1976)}]{PLEBANSKI197698}%
  \BibitemOpen
  \bibfield  {author} {\bibinfo {author} {\bibfnamefont {J.}~\bibnamefont {Plebanski}}\ and\ \bibinfo {author} {\bibfnamefont {M.}~\bibnamefont {Demianski}},\ }\href {\doibase https://doi.org/10.1016/0003-4916(76)90240-2} {\bibfield  {journal} {\bibinfo  {journal} {Annals of Physics}\ }\textbf {\bibinfo {volume} {98}},\ \bibinfo {pages} {98} (\bibinfo {year} {1976})}\BibitemShut {NoStop}%
\bibitem [{\citenamefont {Akcay}\ and\ \citenamefont {Matzner}(2011)}]{Akcay_Matzner_2011}%
  \BibitemOpen
  \bibfield  {author} {\bibinfo {author} {\bibfnamefont {S.}~\bibnamefont {Akcay}}\ and\ \bibinfo {author} {\bibfnamefont {R.}~\bibnamefont {Matzner}},\ }\href {\doibase 10.1088/0264-9381/28/8/085012} {\bibfield  {journal} {\bibinfo  {journal} {Classical and Quantum Gravity}\ }\textbf {\bibinfo {volume} {28}},\ \bibinfo {pages} {085012} (\bibinfo {year} {2011})},\ \bibinfo {note} {arXiv:1011.0479 [gr-qc]}\BibitemShut {NoStop}%
\bibitem [{\citenamefont {Gourgoulhon}(2012)}]{Gourgoulhon2012}%
  \BibitemOpen
  \bibfield  {author} {\bibinfo {author} {\bibfnamefont {E.}~\bibnamefont {Gourgoulhon}},\ }\href@noop {} {\emph {\bibinfo {title} {3+1 Formalism in General Relativity}}},\ Lecture notes in physics\ (\bibinfo  {publisher} {Springer},\ \bibinfo {address} {New York, NY},\ \bibinfo {year} {2012})\BibitemShut {NoStop}%
\bibitem [{\citenamefont {Clough}\ \emph {et~al.}(2022)\citenamefont {Clough}, \citenamefont {Helfer}, \citenamefont {Witek} \emph {et~al.}}]{Clough_2022}%
  \BibitemOpen
  \bibfield  {author} {\bibinfo {author} {\bibfnamefont {K.}~\bibnamefont {Clough}}, \bibinfo {author} {\bibfnamefont {T.}~\bibnamefont {Helfer}}, \bibinfo {author} {\bibfnamefont {H.}~\bibnamefont {Witek}},  \emph {et~al.},\ }\href {\doibase 10.1103/physrevlett.129.151102} {\bibfield  {journal} {\bibinfo  {journal} {Physical Review Letters}\ }\textbf {\bibinfo {volume} {129}} (\bibinfo {year} {2022}),\ 10.1103/physrevlett.129.151102}\BibitemShut {NoStop}%
\bibitem [{\citenamefont {Zilh\~{a}o}\ \emph {et~al.}(2015)\citenamefont {Zilh\~{a}o}, \citenamefont {Witek},\ and\ \citenamefont {Cardoso}}]{Zilh_o_2015}%
  \BibitemOpen
  \bibfield  {author} {\bibinfo {author} {\bibfnamefont {M.}~\bibnamefont {Zilh\~{a}o}}, \bibinfo {author} {\bibfnamefont {H.}~\bibnamefont {Witek}}, \ and\ \bibinfo {author} {\bibfnamefont {V.}~\bibnamefont {Cardoso}},\ }\href {\doibase 10.1088/0264-9381/32/23/234003} {\bibfield  {journal} {\bibinfo  {journal} {Classical and Quantum Gravity}\ }\textbf {\bibinfo {volume} {32}},\ \bibinfo {pages} {234003} (\bibinfo {year} {2015})},\ \bibinfo {note} {arXiv:1505.00797 [gr-qc, physics:hep-th]}\BibitemShut {NoStop}%
\bibitem [{\citenamefont {Fell}\ and\ \citenamefont {Heisenberg}(2024)}]{fell2024grboondi}%
  \BibitemOpen
  \bibfield  {author} {\bibinfo {author} {\bibfnamefont {S.~D.~B.}\ \bibnamefont {Fell}}\ and\ \bibinfo {author} {\bibfnamefont {L.}~\bibnamefont {Heisenberg}},\ }\href@noop {} {\enquote {\bibinfo {title} {Grboondi: A code for evolving generalized proca theories on arbitrary backgrounds},}\ } (\bibinfo {year} {2024}),\ \Eprint {http://arxiv.org/abs/2405.01348} {arXiv:2405.01348 [gr-qc]} \BibitemShut {NoStop}%
\bibitem [{\citenamefont {Andrade}\ \emph {et~al.}(2021)\citenamefont {Andrade}, \citenamefont {Salo}, \citenamefont {Aurrekoetxea} \emph {et~al.}}]{Andrade2021}%
  \BibitemOpen
  \bibfield  {author} {\bibinfo {author} {\bibfnamefont {T.}~\bibnamefont {Andrade}}, \bibinfo {author} {\bibfnamefont {L.~A.}\ \bibnamefont {Salo}}, \bibinfo {author} {\bibfnamefont {J.~C.}\ \bibnamefont {Aurrekoetxea}},  \emph {et~al.},\ }\href {\doibase 10.21105/joss.03703} {\bibfield  {journal} {\bibinfo  {journal} {Journal of Open Source Software}\ }\textbf {\bibinfo {volume} {6}},\ \bibinfo {pages} {3703} (\bibinfo {year} {2021})}\BibitemShut {NoStop}%
\bibitem [{\citenamefont {Alcubierre}(2008)}]{alcnumrel}%
  \BibitemOpen
  \bibfield  {author} {\bibinfo {author} {\bibfnamefont {M.}~\bibnamefont {Alcubierre}},\ }\href {\doibase 10.1093/acprof:oso/9780199205677.001.0001} {\emph {\bibinfo {title} {Introduction to 3+1 Numerical Relativity}}}\ (\bibinfo  {publisher} {Oxford University Press},\ \bibinfo {year} {2008})\BibitemShut {NoStop}%
\bibitem [{\citenamefont {Fell}()}]{Fell_GRBoondi}%
  \BibitemOpen
  \bibfield  {author} {\bibinfo {author} {\bibfnamefont {S.~D.~B.}\ \bibnamefont {Fell}},\ }\href {https://github.com/ShaunFell/GRBoondi} {\enquote {\bibinfo {title} {Grboondi},}\ }\BibitemShut {NoStop}%
\end{thebibliography}%

\end{document}